\documentclass[a4paper,12pt]{article}
\pdfoutput=1

\usepackage[top=30truemm,bottom=30truemm,left=25truemm,right=25truemm]{geometry}

\usepackage[utf8]{inputenc}
\usepackage{graphicx}
\usepackage{amsmath}
\usepackage{amssymb}
\usepackage{hyperref}
\hypersetup{colorlinks,bookmarksopen,bookmarksnumbered,citecolor=Chromegreen,linkcolor=Royalblue,urlcolor=Royalblue,pdfstartview=FitH,linktocpage}
\usepackage{cite}
\usepackage{braket}
\usepackage{bm}
\usepackage{color}
\usepackage{latexsym}
\usepackage{slashed}
\usepackage{hhline}
\usepackage{enumitem}
\usepackage{mathtools}
\usepackage{comment}

\usepackage{ulem}

\numberwithin{equation}{section}

\allowdisplaybreaks

\definecolor{Royalblue}{rgb}{0.,0.28,0.62}
\definecolor{Chromegreen}{rgb}{0.,0.42,0.23}

\newcommand{\nn}{\nonumber}

\newcommand{\del}{\partial}
\newcommand{\ol}{\overline}

\def\Re{\mathop{\mathrm{Re}}}
\def\Im{\mathop{\mathrm{Im}}}
\def\diag{\mathop{\mathrm{diag}}}

\newcommand{\MNS}{\text{MNS}}
\newcommand{\eq}{\text{eq}}
\newcommand{\LL}{L}
\newcommand{\scat}{\text{scat}}

\newcommand{\rt}{\mathrm{t}}

\graphicspath{{./fig/}}

\begin{document}

\begin{titlepage}

\begin{flushright}
{\ttfamily
KUNS-2895
}
\end{flushright}

\vspace{2cm}

\begin{center}

{\Large \bfseries
Leptonic CP asymmetry and Light flavored scalar%
}

\vspace{1cm}

\renewcommand{\thefootnote}{\fnsymbol{footnote}}
{%
\hypersetup{linkcolor=black}
Yoshihiko Abe$^1$\footnote[1]{\ttfamily yabe3@wisc.edu},
Toshimasa Ito$^2$\footnote[2]{\ttfamily toshimasa.i@gauge.scphys.kyoto-u.ac.jp},
Koichi Yoshioka$^2$\footnote[3]{\ttfamily yoshioka@gauge.scphys.kyoto-u.ac.jp}
}%
\vspace{3mm}

{\itshape%
$^1${Department of Physics, University of Wisconsin, Madison, WI 53706, USA}
\\
$^2${Department of Physics, Kyoto University, Kyoto 606-8502, Japan}
}%

\vspace{1cm}

\abstract{
We consider a situation where right-handed neutrinos couple to a light scalar which is possibly a Nambu-Goldstone boson resulting from high-energy symmetry breaking.
Its coupling is typically complex-valued and flavor-dependent.
In this work, we investigate the possibility of the leptonic asymmetry
generation in the Universe from the right-handed neutrino
decay to flavorful light scalar.
Furthermore a new source of asymmetry generation from a single decay process is pointed out, which is characteristic of the present setting.
}

\end{center}
\end{titlepage}

\renewcommand{\thefootnote}{\#\arabic{footnote}}
\setcounter{footnote}{0}
\setcounter{page}{1}

\tableofcontents

\bigskip

\section{Introduction}

The Standard Model (SM) of particle physics and cosmology still have some mysteries,
e.g.,
the nature of dark matter, the source of matter-antimatter asymmetry,
the origin of neutrino masses, and so on.
An attractive idea to realize tiny neutrino masses is the seesaw mechanism
with Majorana right-handed (RH) neutrinos
\cite{Minkowski:1977sc,Yanagida:1979as,GellMann:1980vs}, called 
the type I seesaw. 
The additional Majorana fermions can also be the origin of matter-antimatter
asymmetry via their decay, the scenario called leptogenesis
\cite{Fukugita:1986hr}.

The baryon asymmetry in the present Universe is measured by the cosmic
microwave background observation \cite{Planck:2018vyg} as
\begin{align}
	Y_{\Delta B} = \frac{n_{\Delta B}}{s} = (0.852 \text{ -- } 0.888) 
	\times 10^{-10},
\end{align}
with the number density $n_{\Delta B}$ and the entropy density $s$.
In the leptogenesis scenario, the RH neutrinos generate the lepton asymmetry, which is then converted to the baryon asymmetry \cite{Kuzmin:1985mm} via the sphaleron process \cite{Klinkhamer:1984di}.
In the simplest leptogenesis, the interference between the tree-level and one-loop level RH neutrino two-body decays gives a source of the asymmetry, which is determined by neutrino Yukawa couplings. 
On the other hand, other possibilities for leptogenesis are studied by considering three-body decays of the RH neutrinos \cite{Adhikari:1996mc,Hambye:2001eu,Abdallah:2019tij} and the dark matter extensions of the SM \cite{Dasgupta:2019lha,Borah:2020ivi}.
Refs.~\cite{LeDall:2014too,Alanne:2018brf} also discuss the model with a real singlet scalar whose Higgs portal coupling plays an important role in leptogenesis from heavier RH neutrino decay.

In this paper,
we focus on a possibility of flavorful light scalar for the lepton
asymmetry generation. This is a minimal scalar extension of the type-I
seesaw model by introducing a SM singlet scalar interacting with the RH
neutrinos. This kind of scalar may be motivated by some dynamical
origin of Majorana neutrino mass scale, 
but in this work we do not specify it and investigate the general form
of scalar interaction to the RH neutrinos.
In this setup, we explore the leptogenesis via the RH
neutrino three-body decays with this flavorful scalar field.

The rest parts are organized as follows.
In Sec.~\ref{sec:model},
the model we focus on is introduced and we derive the decay widths and the asymmetry parameters of the two-body decay processes.
In Sec.~\ref{sec:three-body-decay}, the three-body decay widths and the asymmetry parameter are evaluated and we give their approximation formulae.
In Sec.~\ref{sec:leptogenesis},
we discuss the behaviors of the lepton asymmetry and its approximated relic value by solving the Boltzmann equations,
and show the allowed parameter spaces of this model.
We also give some comments on the dynamics of the additional singlet scalar field and its realization in some phenomenological models.
Sec.~\ref{sec:summary} is devoted to the summary of this paper.

\bigskip

\section{Lagrangian and two-body decay}
\label{sec:model}

In this work, we denote the RH neutrinos as $N_i$ in the mass diagonal
basis. They are the Majorana fermions with the Majorana masses $m_{N_i}$. 
An SM gauge singlet scalar $\chi$ is introduced and assumed to have the
coupling to the RH neutrinos. The lagrangian we consider is 
\begin{align}
	\mathcal{L} 
	= \mathcal{L}_{\mathrm{SM}} + \frac{1}{2} \ol{N_i} 
	( i \slashed{\del} - m_{N_i} ) N_i
	- y^\nu_{ij} \tilde{H}^\dagger \ol{N_i} P_L L_j + \mathrm{h.c.}
	- \frac{1}{2} \chi \ol{N_i} 
	\bigl( \xi_{ij} P_R + \xi^*_{ij} P_L \bigr) N_j,	
	\label{eq:Lagrangian}
\end{align}
where $L_j$ mean the SM lepton doublets and $H$ the Higgs one, and
$P_{L/R}$ is the chirality projection operator. 
The coupling constants $\xi_{ij}$ are $i,j$ symmetric and generally complex-valued. 
As discussed in Ref.~\cite{Casas:2001sr}, with the seesaw relation, 
the neutrino Yukawa coupling $y^\nu$ can be generally parameterized as
\begin{align}
	y^\nu = \frac{\sqrt{2} i}{v} \sqrt{M_N} R \sqrt{m_\nu} U^\dagger_\MNS,
\end{align}
with the Majorana mass matrix $M_N = \diag( m_{N_1}, m_{N_2},
m_{N_3})$ and the neutrino mass 
matrix $m_\nu = \diag(m_{\nu_1}, m_{\nu_2}, m_{\nu_3})$, and the
neutrino flavor mixing matrix $U_\MNS$.
$v \approx 246~\mathrm{GeV}$ is the electroweak scale.
For the neutrino masses and mixing angles, we use the experimentally
observed values summarized in Ref.~\cite{ParticleDataGroup:2020ssz}. $R$ is
an arbitrary complex orthogonal matrix describing the degrees of freedom of
neutrino Yukawa couplings which cannot be reached with the seesaw
relation. We parameterize $R$ as
\begin{align}
	R = \begin{pmatrix}
		1 & 0 & 0 \\
		0 & \cos \omega_{23} & \sin \omega_{23} \\
		0 & - \sin \omega_{23} & \cos \omega_{23}
	\end{pmatrix}
	\begin{pmatrix}
		\cos \omega_{13} & 0 & \sin \omega_{13} \\
		0 & 1 & 0 \\
		- \sin \omega_{13} & 0 & \cos \omega_{13}
	\end{pmatrix}
	\begin{pmatrix}
		\cos \omega_{12} & \sin \omega_{12} & 0 \\
		- \sin \omega_{12} & \cos \omega_{12} & 0 \\
		0 & 0 &1
	\end{pmatrix},
\end{align}
introducing the complex angles $\omega_{ij}$ (called the Casas-Ibarra
(CI) parameters in the following).

Through the neutrino Yukawa coupling, 
the RH neutrinos interact with the SM thermal bath.
For the two-body decay of $N_i$ to the SM particles, the partial
widths at tree level are given by
\begin{align}
	\Gamma_{N_i \to L_j H} = 
	\Gamma_{N_i \to \bar{L}_j \bar{H} } = 
	\frac{y^\nu_{ij} y^{\nu \dagger}_{ji}}{16 \pi } m_{N_i},
\end{align}
where $\bar{L}$, $\bar{H}$ means the corresponding anti-particles. We
denote the total width of the two-body decay via neutrino Yukawa coupling as
\begin{align}
	\tilde\Gamma_i = \sum_j \bigl( \Gamma_{N_i \to L_j H} + \Gamma_{N_i \to \bar{L}_j \bar{H}} \bigr).
\end{align}
When $m_{N_2}>m_{N_1}+m_\chi$, 
a heavier mode $N_2$ can also decay to $N_1$ and $\chi$ via the
off-diagonal coupling $\xi_{12}$ and its width is evaluated as 
\begin{align}
	\Gamma_{N_2 \to N_1 \chi} =&\; \frac{m_{N_2}}{16 \pi} 
	\frac{\sqrt{ \lambda(m_{N_2}^2, m_{N_1}^2 , m_\chi^2)}}{m_{N_2}^2}
	\biggl\{
	\Bigl[  \Bigl( 1- \frac{m_{N_1}}{m_{N_2}} \Bigr)^2 
	- \frac{m_\chi^2}{m_{N_2}^2} \Bigr] ( \Im \xi_{12})^2 
	\nn \\
	&\quad + \Bigl[ \Bigl( 1 + \frac{m_{N_1}}{m_{N_2}} \Bigr)^2 -
	\frac{m_\chi^2}{m_{N_2}^2} \Bigr] (\Re \xi_{12})^2 \biggr\} 
	\label{eq:N2N1chi}
\end{align}
with the kinematic function 
$\lambda(x,y,z)= x^2 + y^2 + z^2 - 2 x y - 2 y z - 2 y x$.

As we will see in the following, 
the decay of the lightest $N_1$ is dominated by the two-body decay, 
but for a heavier mode the resonant contribution of three-body decay
is comparable to that of $N_2\to N_1\chi$. Then the total decay widths
$\Gamma_{N_i}$ of the lightest and heavier RH neutrinos are generally written by
\begin{align}
	\Gamma_{N_1} \approx \tilde\Gamma_1,
	\qquad
	\Gamma_{N_2} = \tilde\Gamma_2 + \Gamma_{N_2 \to N_1 \chi} + \Gamma_2,
\end{align}
where $\Gamma_i$ are the three-body decay widths of $N_i$ and will be
described in the next section.

\begin{figure}[t]
	\centering
	\includegraphics[scale=0.34]{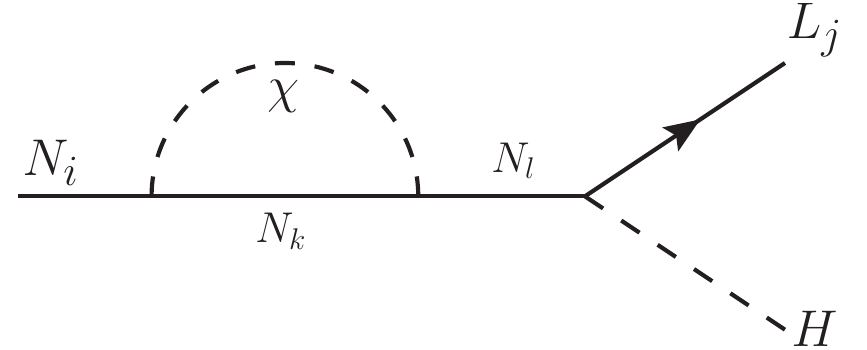}
	\includegraphics[scale=0.34]{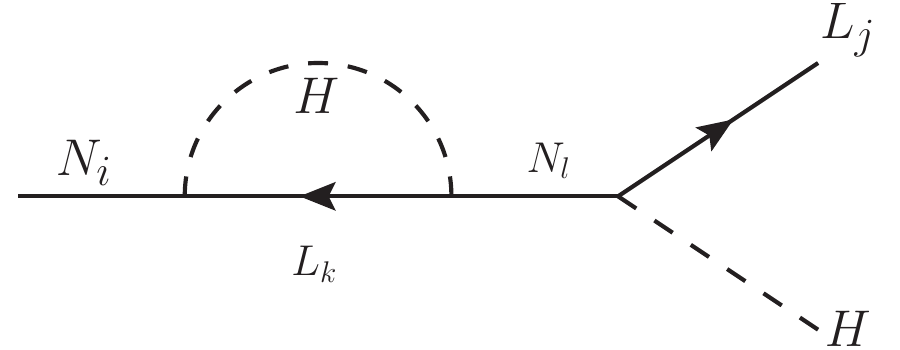}
	\includegraphics[scale=0.34]{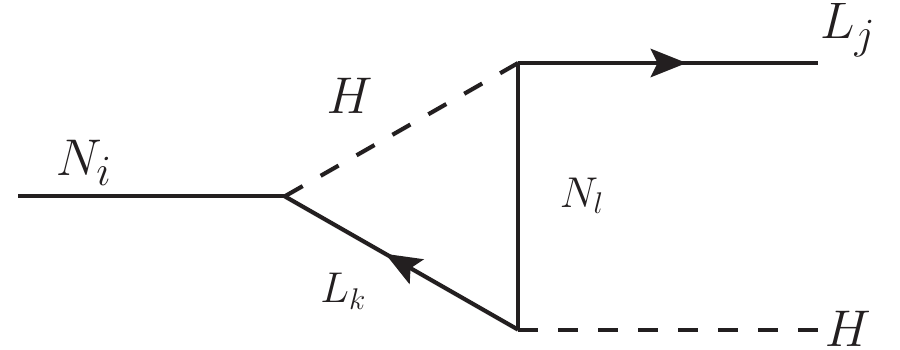}
	\caption{%
		The one-loop two-body decays of a RH neutrino $N_i$ to the SM fields (lepton and Higgs doublets).
	}
	\bigskip
	\label{fig:two-body-decay}
\end{figure}

The CP asymmetry in the ordinary leptogenesis comes
from the interference of the RH neutrino's two-body decays (the tree-level and the right two diagrams in Fig.~\ref{fig:two-body-decay}), where the phases of neutrino Yukawa couplings play an important role. 
The asymmetry parameters $\tilde{\epsilon}^{(LH)}_i$ associated with
this interference is evaluated as
(for a review \cite{Davidson:2008bu}),
\begin{align}
	\tilde{\epsilon}^{(LH)}_{i} = 
	\frac{\Gamma_{N_i \to L H} - \Gamma_{N_i \to \bar{L} \bar{H}}}{\Gamma_{N_i \to L H} + \Gamma_{N_i \to \bar{L} \bar{H}}}
	= \frac{1}{8\pi (y^\nu y^{\nu\dag})_{ii}} \sum_j 
	\Im \bigl[ (y^\nu y^{\nu\dag})_{ji}\bigr]^2
	\mathrm{g}(m_{N_j}^2/ m_{N_i}^2),
	\label{eq:tildeepsilonLH}
\end{align}
where $\mathrm{g} (x)$ in the SM is given by
$ \mathrm{g}(x) = \sqrt{x} \bigl[ \frac{2-x}{1-x} - (1+x) \ln \big(\frac{1+x}{x}\bigr) \bigr]$.
It is noticed that, when $\omega_{ij}$ are real or pure imaginary, 
$\tilde{\epsilon}^{(LH)}_i$ vanish and the
two-body decay via the neutrino Yukawa loop does not generate the CP asymmetry. 
The asymmetry parameters associated with $N \to LH$ is evaluated in the flavor independent manner as Eq.~\eqref{eq:tildeepsilonLH} by summing the flavor index $j$.
The flavor dependence is discussed in \cite{Nardi:2006fx,Abada:2006ea}.

In addition to the SM particle loops,
lighter RH neutrinos and the scalar field $\chi$
give an additional asymmetry in the two-body decay of heavier RH neutrinos.
The left diagram in Fig.~\ref{fig:two-body-decay} shows the relevant
one-loop contribution to the $N_i$ decay. 
The asymmetry reads from the difference between the decay widths to
particles and anti-particles,
\begin{equation}
	\Delta\Gamma_{N\chi}
	= \bigl| \mathcal{M}_{0} + \mathcal{M}_{1} \bigr|^{2}
	-\bigl| \mathcal{M}_{0}^{C} + \mathcal{M}_{1}^{C}
	\bigr|^{2},
\end{equation}
where $\mathcal{M}_0^{(C)}$ denotes the amplitude of
the two-body tree-level decay to the (anti-)particles and
$\mathcal{M}_1^{(C)}$ that of the one-loop diagram including $\chi$,
respectively.
This is evaluated as
\begin{align}
	\Delta \Gamma_{N\chi}
	= \frac{m_{N_i}^4}{4 \pi^2} \sum_{k,l} \frac{1}{m_{N_l}^2-m_{N_i}^2}
	\int_0^1 dx\, \Im \bigl( \ln \Delta^2_{N_k \chi} \bigr)
	\Im \bigl[ f_{li}(x) (y^\nu y^{\nu\dagger})_{il} \bigr],
\end{align}
where 
\begin{align}
	\Delta^2_{N_k \chi} =&\; x(x-1)m_{N_i}^2 + 
	(1-x) m_{N_k}^2 + x m_\chi^2,
	\\
	f_{li} (x) =&\; \biggl(\xi_{lk}^* \xi_{ki} +\xi_{lk} \xi^*_{ki} 
	\frac{m_{N_l}}{m_{N_i}} \biggr) x 
	+ \xi_{lk}^* \xi_{ki}^* \frac{m_{N_k}}{m_{N_i}} 
	+ \xi_{lk} \xi_{ki} \frac{m_{N_k}m_{N_l}}{m_{N_i}^2} .
\end{align}
In order to produce non-trivial asymmetry, the conditions
$i \neq k$ for the interference and 
$\Delta^2_{N_k \chi}<0$
are needed.
The latter condition is satisfied if the internal loop
particles $N_k$ and $\chi$ are lighter than the parent particle $N_i$,
and then only the case of $i=2,~k=l=1$ gives a non-vanishing
asymmetry parameter. As a result, we obtain the $N$-$\chi$ loop 
contribution to asymmetry parameter
\begin{align}\label{TwoBodyEpsilon2}
	\tilde{\epsilon}^{(N\chi)}_{2} =
	\frac{-m_{N_2}}{64 \pi^2 \tilde{\Gamma}_2} \int_{x_-}^{x_+}\! dx\, 
	\Im \bigl[ f_{12}(x) (y^\nu y^{\nu\dagger})_{21} \bigl] 
	\;\approx\,
	\frac{1}{16 \pi (y^{\nu} y^{\nu \dagger})_{22}}
	\Im \bigl[\xi_{11} \xi^*_{12} (y^\nu y^{\nu \dagger})_{12} \bigr],
\end{align}
where $x_\pm$ are the solutions for $\Delta^2_{N_1 \chi}=0$. 
In the latter approximation, we have assumed the hierarchy $m_{N_1}\ll m_{N_2}$.
The result (\ref{TwoBodyEpsilon2}) agrees with the calculation in Ref.~\cite{LeDall:2014too} with the approximation $1-(m_{N_1}/m_{N_2})^2 \approx 1$. 

The total CP asymmetry from two-body decay of the lightest and heavier
RH neutrinos are given by
\begin{equation}
	\tilde{\epsilon}_1 \,=\, \tilde{\epsilon}^{(LH)}_1,
	\qquad
	\tilde{\epsilon}_2 \,=\, \tilde{\epsilon}^{(LH)}_2 + 
	\tilde{\epsilon}^{(N\chi)}_2.
\end{equation}
With general complex phases of neutrino Yukawa couplings, 
$\tilde{\epsilon}^{(LH)}_i$ are not necessarily vanishing 
simultaneously. In the following analysis, 
we simply assume that the CI parameters and the Dirac phase are
real or pure imaginary to 
examine whether the asymmetry can be generated by the effect of the flavorful scalar $\chi$, 
even when the usual CP phases of neutrino Yukawa couplings 
are absent. 
Such structure may indeed be achieved by high-energy 
theoretical assumption, e.g., some symmetry or Yukawa texture. 
In this situation, 
the CP asymmetry from flavor effects also vanishes since the asymmetry nature of two-body decay originates only from the structure of neutrino Yukawa couplings. Further, the condition for the neutrino Yukawa phases implies the vertex loop correction from $\xi$ does not lead to sizable contribution to the lepton asymmetry.

\bigskip

\section{Three-body decay for asymmetry}
\label{sec:three-body-decay}

\begin{figure}[t]
	\centering
	\includegraphics[scale=0.45]{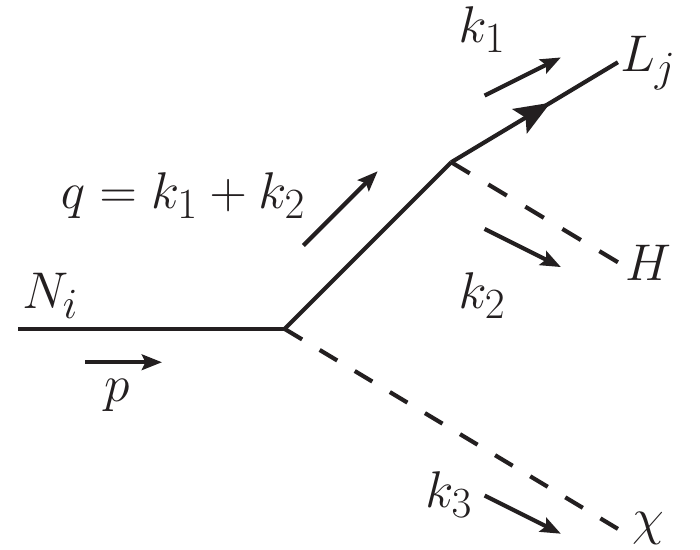}
	\caption{%
		The three-body decay of a RH neutrino $N_i$ at tree level
including the scalar $\chi$.
	}
	\bigskip
	\label{fig:three-body-decays}
\end{figure}

In the following analysis, let us $\chi$ treating as 
a light pseudo-Nambu-Goldstone boson (pNGB) like scalar and the mixing between $\chi$ and the Higgs field is suppressed due to the CP-invariance of the scalar potential. The CP-even partner of pNGB is assumed to be heavy and does not contribute to the asymmetry.  
The detail of the property of $\chi$ and its possible origins are discussed later.

When there exists a scalar $\chi$ and its (pseudo-) scalar coupling to the RH neutrinos $N$, non-trivial CP asymmetry can be generated at the three-body decay $N\to LH\chi$ (Fig.~\ref{fig:three-body-decays}).
Its amplitude is 
\begin{align}
	\mathcal{M}_{N_i \to L_j H \chi} = -
	\sum_k \ol{u_{L_j}}(k_1) y^{\nu \dagger}_{jk} P_R
    \, \frac{1}{\slashed{q} - \bar{m}_{N_k}(q)}
	( \xi_{ki} P_R + \xi^*_{ki} P_L ) u_{N_i}(p),
	\label{eq:3bodyamplitude}
\end{align}
where the momenta $p$, $q$, $k_i$ are shown in the figure. 
The quantity $\bar{m}_{N_k}$ includes the (pole) 
mass and the (finite part of) self energy of $N_k$, which is 
denoted by $\Sigma_{N_k}$.  
While the real part of $\Sigma_{N_k}$ leads to 
the energy-dependent loop corrections apart from the pole, 
we drop them in the present analysis since these small factors do 
not largely change the qualitative property of leptogenesis.  
On the other hand, the imaginary part of $\Sigma_{N_k}$ is important for the CP asymmetry, and is given by the decay 
width of $N_k$ around the pole.
Summing up the final charged-lepton flavor and gauge charge, the three-body decay width becomes
\begin{align}
	\Gamma_{N_i \to L H \chi} =& \sum_{k,\,l} 
	\frac{(y^\nu y^{\nu \dagger})_{lk}}{256 \pi^3 m_{N_i}^3}
	\int_{0}^{(m_{N_i} - m_\chi)^2}\!\!\!dm_{12}^2 
	\int_{[m_{23}^2]_-}^{[m_{23}^2]_+} dm_{23}^2\,
	\frac{1}{(m_{12}^2 - \bar{m}_{N_l}^{*2})(m_{12}^2 - \bar{m}_{N_k}^2)}
	\nn \\
	&\;
	\times \Bigl[ \xi_{il} \xi^*_{ki} m_{12}^2 ( m_{12}^2 + m_{23}^2 - m_\chi^2)
	+ (\xi_{il}\xi_{ki} \bar{m}_{N_k} + 
	\xi^*_{il}\xi^*_{ki} \bar{m}_{N_l}^* ) m_{N_i} m_{12}^2
	\nn \\
	& \qquad
	+ \xi^*_{il} \xi_{ki} \bar{m}_{N_l}^* \bar{m}_{N_k} ( m_{N_i}^2 - m_{23}^2)
	\Bigr].
	\label{eq:Gamma_N_itoLHchi}
\end{align}
The invariant mass parameters are defined by
$m_{12}^2 = (k_1 + k_2)^2$, $m_{23}^2 = (k_2 + k_3)^2$, and the
maximum and minimum values of $m_{23}^2$ with a fixed $s$ are given 
by \cite{Kumar:1969jjy,ParticleDataGroup:2020ssz}
\begin{align}
	[m_{23}^2]_{\pm} = \frac{1}{2} \left[
	m_{N_i}^2 + m_\chi^2 - m_{12}^2 \pm 
	\sqrt{\lambda(m_{N_i}^2 , m_\chi^2, m_{12}^2)} \,\right].
\end{align}
The three-body decay to the anti-particles 
$\Gamma_{N_i \to \bar{L} \bar{H} \chi}$ is evaluated in the same
manner. The total decay width and the CP asymmetry parameter 
for the three-body decay are defined by
\begin{align}
	\Gamma_i = \Gamma_{N_i \to L H \chi} + 
	\Gamma_{N_i \to \bar{L} \bar{H} \chi},
	\qquad
	\epsilon_i = \frac{ \Gamma_{N_i \to L H \chi} - \Gamma_{N_i \to \bar{L} \bar{H} \chi} }{ \Gamma_{N_i \to L H \chi} + \Gamma_{N_i \to \bar{L} \bar{H} \chi} }.
\end{align}
In the following, we will discuss dominant contributions to the widths
and asymmetry parameters and their approximate forms.

\medskip

\subsection{The lightest mode}
\label{sec:lightest-mode}

Among the intermediate states $N_k$, the lightest one $N_1$ generally
gives the dominant contribution to the $N_1$ three-body decay, i.e.,
the $i=k=l=1$ part in \eqref{eq:Gamma_N_itoLHchi}. For a typical
hierarchy of mass parameters, $m_\chi\ll m_{N_1}\ll m_{N_2}$, the $N_1$
decay width reduces to
\begin{align}
	\Gamma_1^{(11)} \approx 
	\frac{m_{N_1}}{64\pi^3} (y^\nu y^{\nu\dagger})_{11}
	\big[ |\xi_{11}|^2 +\Re (\xi^2_{11})\big]
	\ln \Bigl( \frac{m_{N_1}}{m_\chi} \Bigr),
\end{align}
which is mainly determined by the diagonal coupling
$\xi_{11}$. Compared with the two-body decay $N_1\to LH$, 
the width is multiplied 
by a factor $\sim(\xi_{11})^2/4\pi\,\ln(m_{N_1}/m_\chi)$, which 
can be of order one. If $\xi_{11}$ is negligibly small, the width is
instead given by the $k=l=2$ part,
\begin{align}
	\Gamma_1^{(22)} \approx \frac{m_{N_1}^3}{384\pi^3 m_{N_2}^2} 
	(y^\nu y^{\nu\dagger})_{22} |\xi_{12}|^2.
\end{align}

The CP asymmetry, the difference between the decay widths to particles and anti-particles, is usually described by the quantum interference of processes with different intermediate states. 
For the current three-body decay, such interference occurs from the diagrams with different intermediate neutrinos $N_{k,l}$ $(k\neq l)$, that is, the cross terms in the amplitude squared. It is important that, in addition to this usual asymmetry, the present model induces CP asymmetry from a single diagram with one intermediate state.
For example, evaluating Eq.~\eqref{eq:Gamma_N_itoLHchi} with the energy dependent part of the propagator makes us to find that the $N_1$ three-body decay with the intermediate $N_1$ leads to  
a nonzero CP asymmetry parameter
\begin{align}
	\epsilon_1^{(11)} \approx 
	\,\frac{\gamma_{N_1}}{2m_{N_1}}\,
	\frac{\Im (\xi_{11}^2)}{|\xi_{11}|^2 +\Re (\xi^2_{11})},
	\label{eq:N1eps11}
\end{align}
where $\gamma_{N_1}$ is almost the same to the 
decay width of $N_1$ up to a factor of 1 in the limit that 
the initial state is much heavier than the others as we will assume in the numerical analysis.
This is the asymmetry irrelevant to the $N_2$ physics at the 
leading order of $m_\chi\ll m_{N_1}\ll m_{N_2}$. The
exact form of asymmetry parameter is found in Appendix \ref{sec:cpasym}.

\begin{figure}[t]
	\centering
	\includegraphics[scale=0.42]{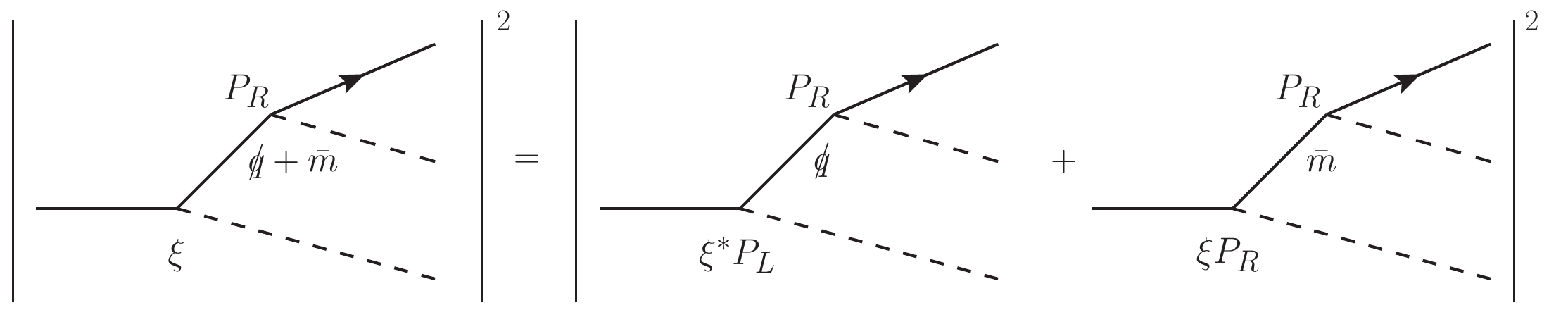}
	\caption{CP asymmetry (cross term) from a single amplitude squared.}
	\bigskip
	\label{fig:cpsingle}
\end{figure}

The non-vanishing ``interference'' from a single 
diagram \eqref{eq:N1eps11} is a characteristic feature of the present
decay mode and its origin is understood by the amplitude form. The
three-body decay amplitude \eqref{eq:3bodyamplitude} contains the factor 
$P_R(\slashed{q}+\bar{m}_N)(\xi P_R+\xi^*P_L)$ 
between the initial and final spinor wavefunctions.
The first projection $P_R$ comes from the Yukawa (chirality-violating)
vertex, the second factor $(\slashed{q}+\bar{m}_N)$ the propagator of
heavy unstable fermion, 
and the third one $(\xi P_R+\xi^*P_L)$ means the Majorana fermion
vertex. The chirality structure implies this factor is divided into
two pieces $\bar{m}_N\xi$ and $\slashed{q}\xi^*$, and hence the
squared amplitude generally contains a cross term of these two
pieces (Fig.~\ref{fig:cpsingle}). 
Further the corresponding decay width to anti-particles is
obtained by the replacement of couplings 
$\xi\leftrightarrow\xi^*$ (and $y^\nu\leftrightarrow y^{\nu*}$, 
$P_R\leftrightarrow P_L$). In the end, the interference of these two
pieces leads to the CP asymmetry proportional 
to $\Im(\xi^2)\Im(\bar{m})$, appearing in 
the numerator of \eqref{eq:N1eps11}. We thus find that this CP
asymmetry from a single decay process is generated in the presence 
of a chiral vertex, a unstable intermediate state, and a complex
decay coupling. The three-body decay of a RH neutrino to the
SM particles plus scalar is an interesting realization of all these
criteria satisfied.

The interference of different intermediate states also 
contributes to the CP asymmetry as usual, and its approximate
form is found
\begin{align}
	\epsilon_1^{(12)} \approx
	\frac{m_{N_1}\Gamma_{N_2}}{8m_{N_2}^2} 
	\frac{\Im\bigl[ (y^\nu y^{\nu \dagger})_{12} \xi_{12}
	(2\xi_{11}+3\xi_{11}^*) 
	\bigr]}{(y^\nu y^{\nu \dagger})_{11}\big[|\xi_{11}|^2 +\Re (\xi^2_{11})\big]
	\ln \bigl( \frac{m_{N_1}}{m_\chi} \bigr)} \,.
	\label{eq:N1eps12}
\end{align}
Which contribution \eqref{eq:N1eps11} or \eqref{eq:N1eps12} is the
dominant CP asymmetry depends on the model parameters. 

The $k=l=2$ part in the $N_1$ decay width only gives a subdominant CP
asymmetry in almost case due to the large $m_{N_2}$ suppression.

\medskip

\subsection{Heavier modes}

A heavier RH neutrino than $N_1$ can also
generate CP asymmetry at its three-body decay. A lighter intermediate state,
e.g., $N_1$ meets the resonance around its mass $m_{12}^2\sim m_{N_1}^2$ and
then the amplitude is largely enhanced. There are two types of
enhancement in the decay width \eqref{eq:Gamma_N_itoLHchi}, the
$k=l=1$ and cross terms. The enhanced contributions to the decay width
are evaluated with the narrow width approximation. For the $k=l=1$
part, we obtain 
\begin{align}
	\Gamma_2^{(11)} =&~
	\frac{m_{N_1}}{128\pi^3m_{N_2}^3\Gamma_{N_1}} (y^\nu y^{\nu\dagger})_{11}
	\sqrt{\lambda(m_{N_1}^2,m_{N_2}^2,m_\chi^2)}  \nn \\
	& \quad \times\big[|\xi_{12}|^2(m_{N_1}^2+m_{N_2}^2-m_\chi^2) +2\Re
	(\xi^2_{12})m_{N_1}m_{N_2} \big] ,
	\label{eq:N2width11}
\end{align}
where we have neglected $\mathcal{O}((\Gamma_{N_1})^2)$ terms.
In the limit $m_\chi\ll m_{N_1}\ll m_{N_2}$ and
$\Gamma_{N_1} \approx \tilde\Gamma_1$, the 
$N_1$ resonant contribution \eqref{eq:N2width11} is equal 
to $\frac{1}{2}\Gamma_{N_2\to N_1\chi}$. (The prefactor $\frac{1}{2}$
implies $\tilde\Gamma_1$ contains the decays both to particles and
anti-particles.) We also find the cross-term 
contribution is relatively suppressed than the $k=l=1$ one. 
The enhanced on-shell decay \eqref{eq:N2width11} is controlled by the 
off-diagonal coupling $\xi_{12}$. When $\xi_{12}$ is negligibly small,
the $N_2$ three-body decay width is dominantly given by the
non-resonant $k=l=2$ part and its approximate form is
\begin{align}
	\Gamma_2^{(22)} \approx 
 	\frac{m_{N_2}}{64\pi^3} (y^\nu y^{\nu\dagger})_{22}
  	\big[|\xi_{22}|^2 +\Re (\xi^2_{22})\big]
  	\ln \Bigl( \frac{m_{N_2}}{m_\chi} \Bigr) .
\end{align}

The CP asymmetry of $N_2$ three-body decay is also generated by the
resonant and non-resonant parts. The asymmetry from the $N_1$
resonant part comes from the last term in \eqref{eq:N2width11} and reads
\begin{align}
  	\epsilon_2^{(11)} =
  	\frac{m_{N_1}}{128\pi^2m_{N_2}^2\Gamma_2} (y^\nu y^{\nu\dagger})_{11}
  	\sqrt{\lambda(m_{N_1}^2,m_{N_2}^2,m_\chi^2)} \Im(\xi^2_{12}).
  	\label{eq:N2eps11}
\end{align}
The usual interference of different diagrams also contributes to the
asymmetry and is approximately found 
\begin{align}
  	\epsilon_2^{(12)} \approx
  	\frac{-m_{N_1}}{128\pi^2\Gamma_2} \Im\big[(y^\nu y^{\nu\dagger})_{21}
  	\xi_{22}^*\xi_{12}\big]
  	\label{eq:N2eps12}
\end{align}
in the limit $m_{N_1}\ll m_{N_2}$. This usual cross-term part 
gives a tiny contribution to the decay width as mentioned above, but a
possibly large one to the CP asymmetry parameter. 
The exact form of asymmetry parameters from the resonant contribution
is found in Appendix \ref{sec:cpasym}.

\begin{figure}[t]
	\centering
	\includegraphics[width=0.46\textwidth]{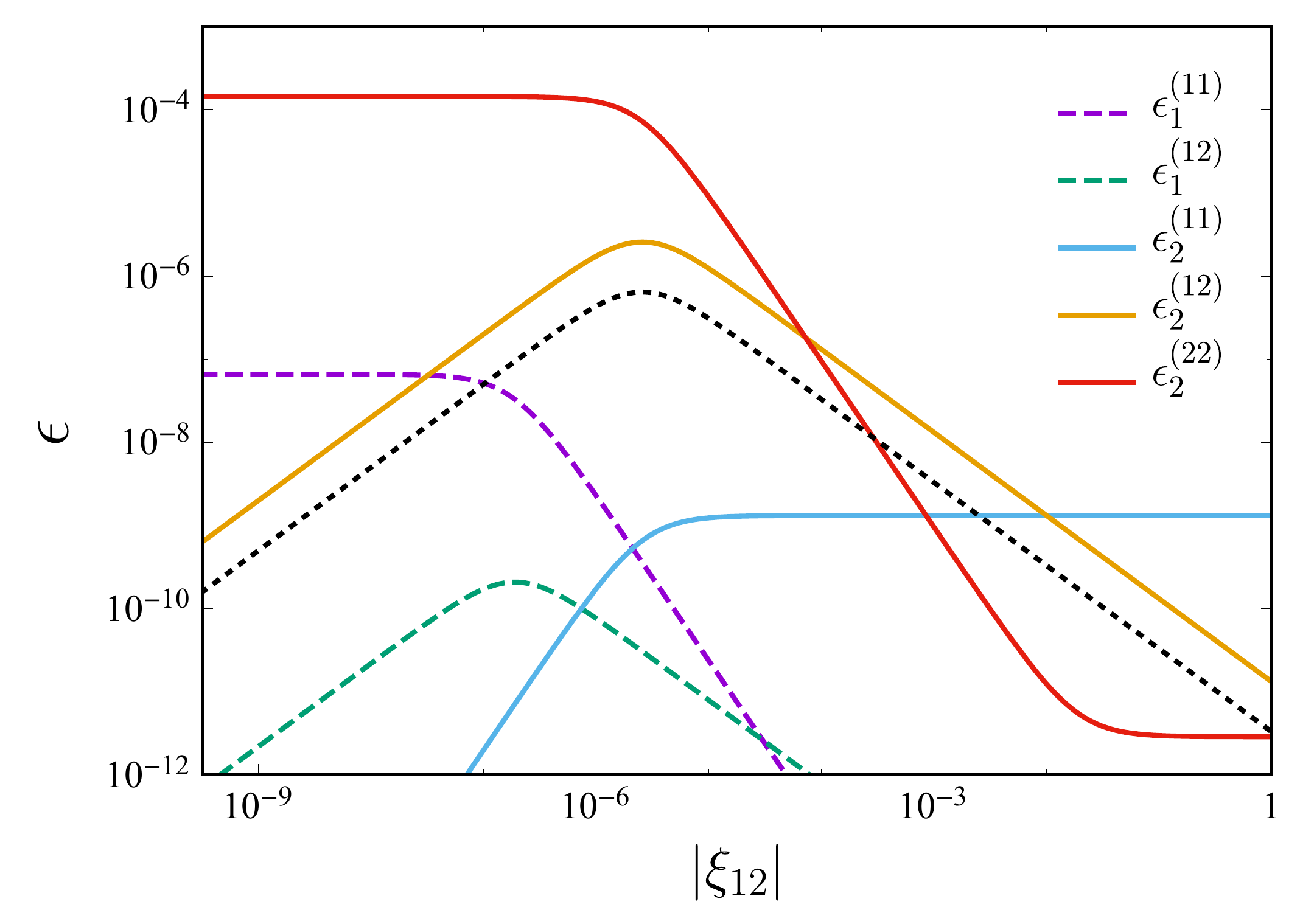}
	\qquad
	\includegraphics[width=0.46\textwidth]{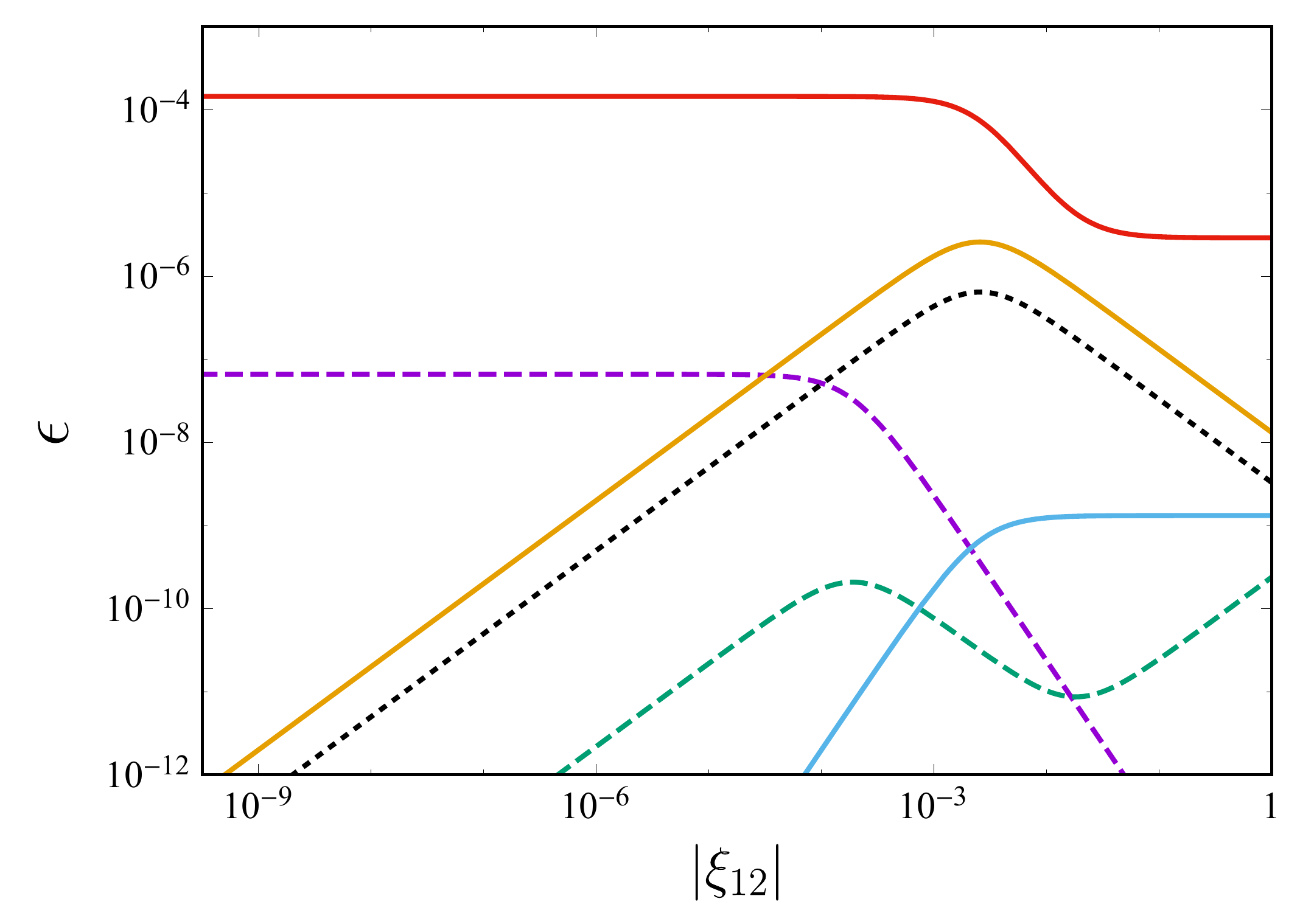}
	\caption{%
		A typical behavior of various asymmetry parameters. The
		approximate formulae of these parameters are given in the text.
		(Left) $|\xi_{11}|=10^{-5}$ and $|\xi_{22}|=10^{-3}$.
		(Right) $|\xi_{11}|=10^{-2}$ and $|\xi_{22}|=1$.
		In these figures, the
		neutrino Yukawa couplings are roughly replaced by the mass eigenvalues
		of neutrinos as $(y^\nu y^{\nu \dagger})_{ij}\sim 
		(m_{N_i}m_{N_j}m_{\nu_i}m_{\nu_j})^{1/2}/v^2$, and the complex phases
		of couplings are assumed such that the asymmetry parameters take their maximal values. 
        For comparison, the relative size of asymmetry parameter for the two-body
        decay $\tilde{\epsilon}_2(\tilde{\Gamma}_2/\Gamma_2)$ is shown by the black dotted line.
    }
	\bigskip
	\label{fig:epsilons}
\end{figure}

Similar to the $N_1$ decay, there is the CP asymmetry from a single diagram 
where the unstable $N_2$ is the intermediate state.
In this case, the difference of the decay widths to particles and anti-particles is the same form as in the $N_1$ decay, but the total three-body decay width has several possibilities as described above.
The resultant asymmetry parameter from the $N_2$-intermediate diagram becomes
\begin{align}
  	\epsilon_2^{(22)} \approx 
  	\frac{\gamma_{N_2}}{128 \pi^3\Gamma_2} (y^\nu y^{\nu \dagger})_{22}
  	\Im (\xi_{22}^2)  \ln \Bigl(\frac{m_{N_2}}{m_\chi} \Bigr),
  	\label{eq:N2eps22}
\end{align}
where $\gamma_{N_2}$ is defined in the same way 
as $\gamma_{N_1}$ in Eq.~\eqref{eq:N1eps11}.
Among the above 3 types of $\epsilon_2$, the leading
contribution is generally given by the non-resonant part
\eqref{eq:N2eps22} except for a tiny $\xi_{22}$ (see Fig.~\ref{fig:epsilons}).
We thus find for the $N_2$ three-body decay that the width is
determined by the diagonal $k=l=1$ or $k=l=2$ part, and the CP
asymmetry is governed by the non-resonant $k=l=2$ part.

\bigskip

\section{Lepton asymmetry with $\chi$ scalar}
\label{sec:leptogenesis}

\subsection{Boltzmann equations}

In this section, solving the Boltzmann equations in the present
system, we discuss the time evolution of yields and
the washout effect of asymmetry. 
When the temperature of the Universe is denoted by $T$, 
the Hubble parameter $H$ and the entropy density $s$
in the radiation-dominated Universe are defined by 
\begin{align}
	H (T) = \sqrt{ \frac{\pi^2}{90} g_*} \frac{T^2}{M_P},
	\qquad
	s(T) = \frac{2 \pi^2}{45} g^S_* T^3,
\end{align}
where $M_P = 2.44 \times 10^{18}~\mathrm{GeV}$ is the reduced Planck scale,
$g_*$ ($g_*^S$) denotes the effective degrees of freedom for the
energy (entropy) density. 
The yield of particle $i$ is defined by $Y_i = n_i / s$ with the number density $n_i$, and $Y^\eq_i$ is that of the thermal equilibrium.

In the following analysis, we apply for simplicity the single flavor
approximation to the lepton asymmetry, 
and write the yields of leptons and anti-leptons as
\begin{align}
	Y_L = Y^\eq_\LL + \frac{1}{2} Y_{\Delta L},
	\qquad
	Y_{\bar{L}} = Y^\eq_\LL - \frac{1}{2} Y_{\Delta L},
\end{align}
where $Y_{\Delta L}$ denotes the yield of the lepton asymmetry.
If one would like to include flavor-dependent effects, a further detailed analysis with, e.g., the density matrix formalism, is needed \cite{Granelli:2021fyc} but that is beyond the purpose of this work. 
We also drop theral effects in evaluating the Boltzmann equations which could lead some numerical changes.
The SM particles including the leptons and the Higgs field are assumed to be in the thermal bath. 
In this setup, the Boltzmann equations needed for the lepton asymmetry are given by
\begin{align}
	H x \frac{dY_{N_1}}{dx} \approx &\;
	\frac{K_1 (m_{N_1}/T)}{K_2( m_{N_1}/T)} Y^\eq_{N_1}  \Biggl[ \tilde\Gamma_1 \biggl( 1 - \frac{Y_{N_1}}{Y^\eq_{N_1}} \biggr)
	- \frac{1}{2} \tilde\Gamma_1 \tilde\epsilon_1 \frac{Y_{\Delta L}}{Y^\eq_\LL}
	+ \Gamma_1 \biggl( - \frac{Y_{N_1}}{Y^\eq_{N_1}} + \frac{Y_\chi}{Y^\eq_\chi} \biggr)
	\nn \\
	&\; - \frac{1}{2} \Gamma_1 \epsilon_1 \frac{Y_\chi}{Y^\eq_\chi} \frac{Y_{\Delta L}}{Y^\eq_\LL} \Biggr]
	\nn \\
	& \;
	+ \frac{K_1 (m_{N_2} /T)}{K_2(m_{N_2}/ T)} Y^\eq_{N_2} \Gamma_{N_2 \to N_1 \chi} \biggl( \frac{Y_{N_2}}{Y_{N_2}^\eq} - \frac{Y_{N_1}}{Y^\eq_{N_1}} \frac{Y_\chi}{Y^\eq_{\chi}} \biggr) +C_\scat ,
	\label{eq:BE-N_1-sfa}
	\\
 	H x \frac{ d Y_{N_2}}{dx} \approx &\;
 	\frac{K_1(m_{N_2}/T)}{K_2(m_{N_2}/T)} Y^\eq_{N_2} \Biggl[ \tilde\Gamma_2 \biggl( 1 - \frac{Y_{N_2}}{Y^\eq_{N_2}} \biggr)
    - \frac{1}{2} \tilde\Gamma_2 \tilde\epsilon_2 \frac{Y_{\Delta L}}{Y^\eq_\LL}
 	+ \Gamma_2 \biggl( - \frac{Y_{N_2}}{Y^\eq_{N_2}} + \frac{Y_\chi}{Y^\eq_\chi} \biggr)
 	\nn \\
 	&\; - \frac{1}{2} \Gamma_2 \epsilon_2 \frac{Y_\chi}{Y^\eq_\chi} \frac{Y_{\Delta L}}{Y^\eq_\LL}
 	- \Gamma_{N_2 \to N_1 \chi} \biggl( \frac{Y_{N_2}}{Y^\eq_{N_2}} - \frac{Y_{N_1}}{Y^\eq_{N_1}} \frac{Y_\chi}{Y_\chi^\eq} \biggr) \Biggr] +C_\scat ,
 	\label{eq:BE-N_2-sfa}
 	\\
 	H x \frac{d Y_{\Delta L}}{dx} \approx &\;
 	\sum_i
 	\frac{K_1 (m_{N_i}/T)}{K_2 (m_{N_i}/T)} Y^\eq_{N_i}
 	\Biggl[ \tilde\Gamma_i \tilde\epsilon_i \biggl( \frac{Y_{N_i}}{Y^\eq_{N_i}} - 1 \biggr)
 	- \frac{1}{2} \tilde\Gamma_i \frac{Y_{\Delta L}}{Y^\eq_\LL}
 	\nn \\
 	&\; + \Gamma_i \epsilon_i \biggl( \frac{Y_{N_i}}{Y^\eq_{N_i}} - \frac{Y_\chi}{Y^\eq_\chi} \biggr)
 	- \frac{1}{2} \Gamma_i \frac{Y_\chi}{Y^\eq_\chi} \frac{Y_{\Delta L}}{Y^\eq_\LL}
 	\Biggr] + C_\scat ,
 	\label{eq:BE-DeltaL-sfa}
 	\\
 	H x \frac{d Y_\chi}{dx} \approx &\;
 	\sum_i \frac{K_1 (m_{N_i}/T)}{K_2(m_{N_i}/T)} Y^\eq_{N_i} \Biggl[
 	\Gamma_i \biggl( \frac{Y_{N_i}}{Y^\eq_{N_i}} - \frac{Y_\chi}{Y_\chi^\eq} \biggr)
 	+ \frac{1}{2} \Gamma_i \epsilon_i \frac{Y_\chi}{Y_\chi^\eq} \frac{Y_{\Delta L}}{Y^\eq_\LL} \Biggr] + C_\scat
 	\nn \\
 	&\; + Y^\eq_{N_2} \frac{K_1 (m_{N_2}/T)}{K_2 (m_{N_2}/T)} \Gamma_{N_2 \to N_1 \chi} \biggl( \frac{Y_{N_2}}{Y^\eq_{N_2}} - \frac{Y_{N_1}}{Y^\eq_{N_1}} \frac{Y_\chi}{Y^\eq_\chi} \biggr) + C_\scat ,
 	\label{eq:BE-chi-sfa}
\end{align}
where the dimensionless parameter $x$ is introduced by $x = m_{N_1}/T$, $K_n(z)$ denotes the modified Bessel function of second kind.
The $C_\scat$ parts are the collision terms from the scattering with the on-shell contributions removed.
For the three-body decay process, the on-shell contributions of RH neutrinos  are deducted in order to avoid the double counting when deriving the Boltzmann equations \eqref{eq:BE-N_1-sfa}--\eqref{eq:BE-chi-sfa}. 
\begin{table}[t]
	\centering
	\begin{tabular}{|c||c|c|} \hline
 		& $N_2$ decay dominant & $N_1$ decay dominant \\
 		\hhline{|=#=|=|}
 		weak washout & $\mathcal{Y}^{\mathrm{FI}}_{\Delta L}(z_1)$ & $Y^{\mathrm{FI}}_{\Delta L}(\infty)$ \\
 		\hline
 		strong washout & -----  & $\mathcal{Y}^{\mathrm{WO}}_{\Delta L}(\infty)$
 		\\ \hline
	\end{tabular}
	\caption{
        Four possible patterns of dominant contributions to the lepton asymmetry.
    }
	\bigskip
	\label{tab:four-main-processes}
\end{table}

\begin{figure}[t]
	\centering
	\includegraphics[width=0.49\textwidth]{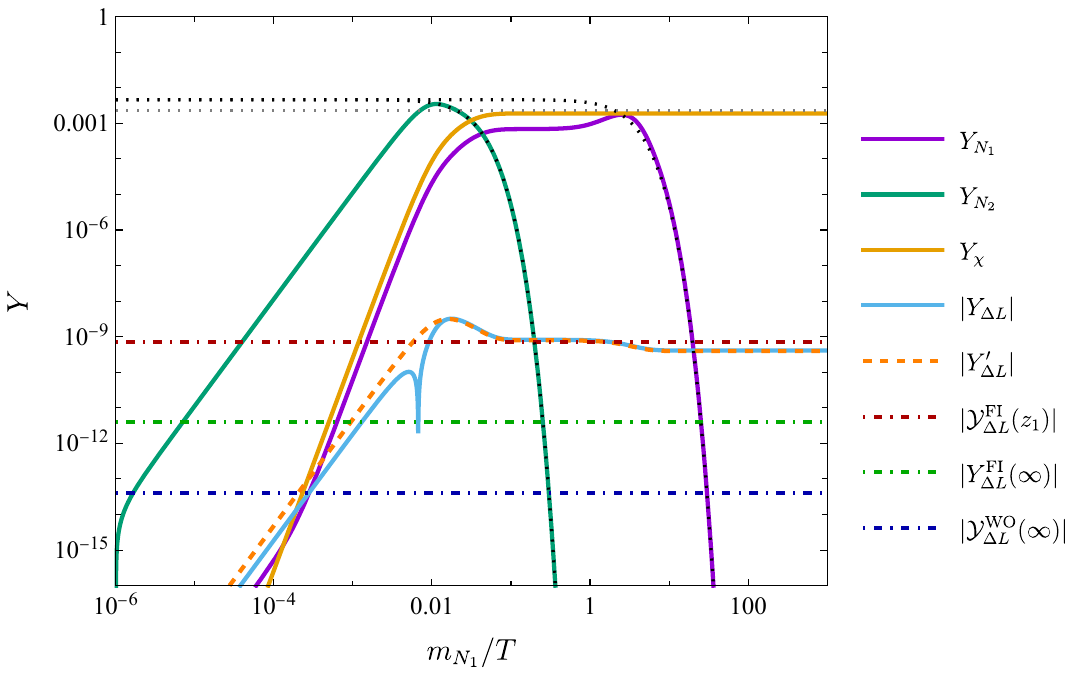}
	\includegraphics[width=0.49\textwidth]{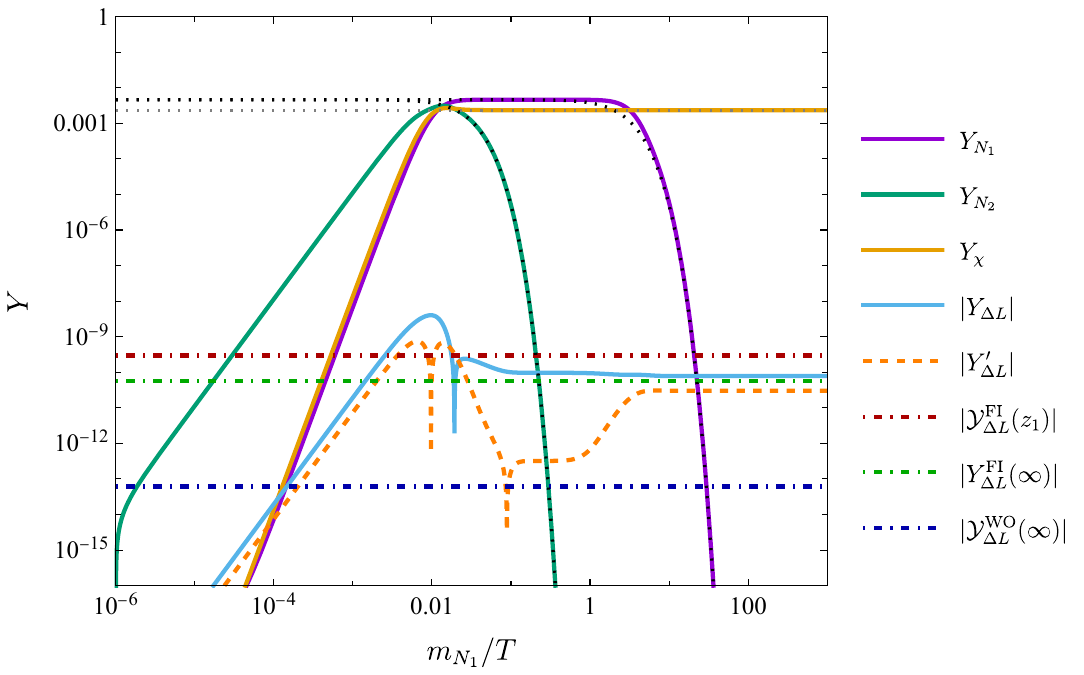}
	\\
	\includegraphics[width=0.49\textwidth]{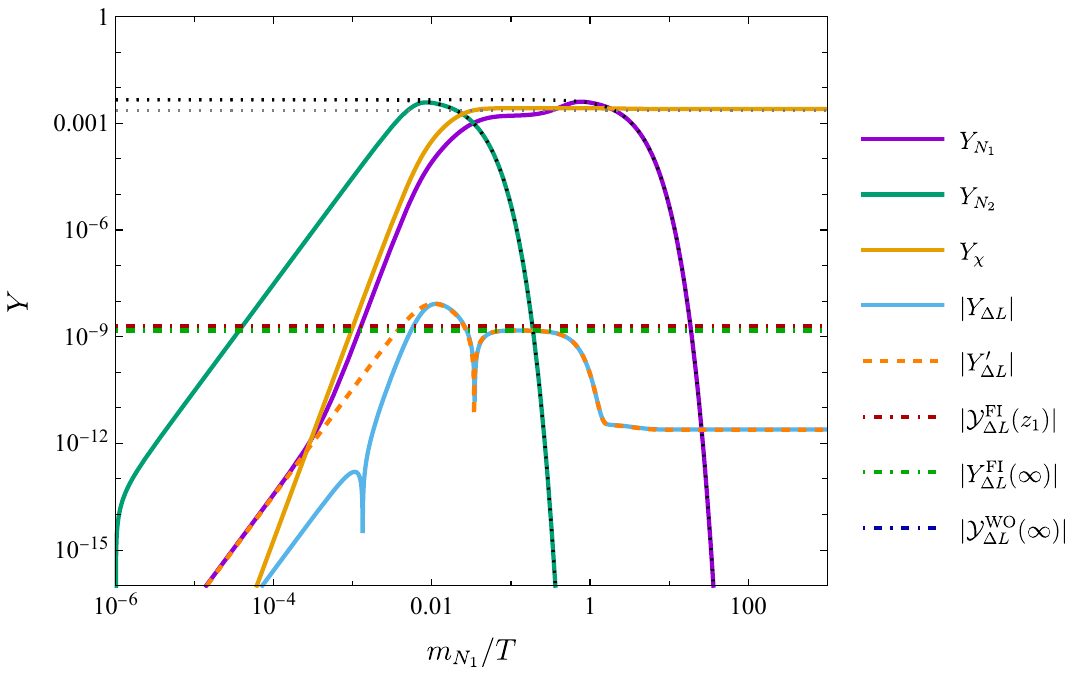}
	\includegraphics[width=0.49\textwidth]{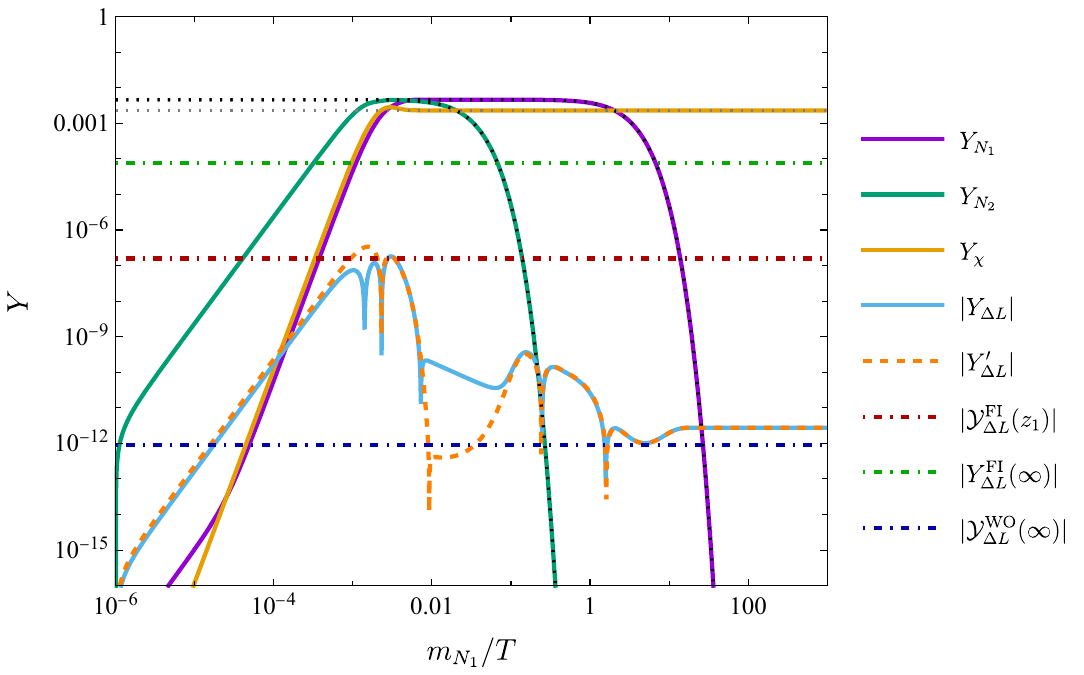}
	\caption{%
		The time evolution of the yields. In the top panels,
		the washout effect is weak and the asymmetry is produced by the freeze-in mechanism of the RH neutrino decays.
		The washout works well in the bottom panels.
		In the left panels, the $N_2$ decay is the dominant production
		channels and $N_1$ dominant in the right panels. 
		The solid lines are numerical solutions of the Boltzmann equations
		[Eqs.~\eqref{eq:BE-N_1-sfa}--\eqref{eq:BE-chi-sfa}]. 
		The orange dashed line is evaluated by solving the Boltzmann equations of $N_1$, $\chi$ and $\Delta L$ with $Y_{N_2} = Y^\eq_{N_2}$.
		The dot-dashed dark red, dark green and dark blue lines are given by Eqs.~\eqref{eq:mathcalY-FI}, \eqref{eq:YFI(infty)} and \eqref{eq:mathcalY-WO}, respectively.
	}
	\bigskip
	\label{fig:sol-BE}
\end{figure}

The RH neutrinos are first generated by their interactions to the thermal bath, and the lepton asymmetry and the scalar $\chi$ are produced via the RH neutrino decays.
While the total amounts of the lepton asymmetry and $\chi$ are related, the washout (inverse decay) effect via the neutrino Yukawa couplings deduces only the asymmetry. 
The Boltzmann equation for the lepton asymmetry approximately becomes
\begin{align*}
	\frac{d Y_{\Delta L}}{dx} \approx \sum_i \frac{1}{H x}
	\frac{K_1 (m_{N_i} x / m_{N_1})}{K_2 (m_{N_i} x / m_{N_1})}
	\biggl[
	    \Gamma_i \epsilon_i \biggl( Y_{N_i} - \frac{Y^\eq_{N_i} Y_\chi}{Y^\eq_\chi} \biggr)
	    - \frac{1}{2} \biggl( \tilde{\Gamma}_i +  \Gamma_i \frac{Y_\chi}{Y^\eq_\chi} \biggr) \frac{Y^\eq_{N_i}}{Y^\eq_\LL} Y_{\Delta L}
    \biggr],
\end{align*}
where the $\tilde{\epsilon}_i$ terms have been dropped since we are interested in the asymmetry produced by the three-body decay process.
The first term of RHS produces the asymmetry via the $N_i$ decay and the second term denotes the washout term. Depending on the $N_1$ or $N_2$ decay process being dominant to the production and the washout effect being strong or weak, there are four possibilities of the main lepton asymmetry as listed in Table~\ref{tab:four-main-processes}. 
We also show in Fig.~\ref{fig:sol-BE} the typical time evolution of the
yields corresponding to these four patterns, respectively.
The top panels in Fig.~\ref{fig:sol-BE} are the results for the weak washout, and the bottom panels are for the strong one. Further, the $N_2$ decay process is the main contribution to the asymmetry in the left panels in Fig.~\ref{fig:sol-BE} and the $N_1$ process is dominant in the right panels.
Roughly speaking, the relic of lepton asymmetry in the bottom line is given by that in the top line times the washout suppression factor.
The dips in Fig.~\ref{fig:sol-BE} come from the sign flip in the right-hand side of Eq.~\eqref{eq:BE-DeltaL-sfa} before and after the particles getting into the thermal equilibrium.
This behavior depends on the magnitude of the coupling, and we find more dips in the strong washout regime.

In the case that the freeze-in production from $N_2$ is dominant 
(top-left panel), we have to take the washout effect into account (for
the detail, see Appendix~\ref{sec:formula}). 
The lepton asymmetry of this case is evaluated by
\begin{align}
 \mathcal{Y}^{\mathrm{FI}}_{\Delta L} (z_1) = \int_{0}^{z_1} dx' \, \mathcal{F}_2 (x')
 \exp \biggl[ - \int_{x'}^{z_1} dx''\, W_2(x'') \biggr],
 \label{eq:mathcalY-FI}
\end{align}
where
\begin{align}
 W_2 &\coloneqq \frac{1}{2 H x} \frac{K_1 ( m_{N_2} x / m_{N_1})}{K_2 ( m_{N_2} x / m_{N_1})}
 \biggl( \tilde{\Gamma}_2 + \Gamma_2 \frac{\bm{Y}_\chi}{Y^\eq_\chi} \biggr)
 \frac{Y^\eq_{N_2}}{Y^\eq_\LL}
 + \frac{1}{2 H x} \frac{K_1 (x)}{K_2(x)}
 \biggl( \tilde{\Gamma}_1 + \Gamma_1 \frac{\bm{Y}_\chi}{Y^\eq_\chi} \biggr)
 \frac{Y^\eq_{N_2}}{Y^\eq_\LL},
 \\
 \mathcal{F}_2 =&\;
 \frac{1}{H x}  \frac{K_1 (m_{N_2}x /m_{N_1})}{K_2(m_{N_2}x/ m_{N_1})}
 \Gamma_2 \epsilon_2 \biggl( \bm{Y}_{N_2} - \frac{Y^\eq_{N_2}}{Y^\eq_\chi}
 \bm{Y}_\chi \biggr)
 + \frac{1}{H x} \frac{K_1(x)}{K_2(x)} \Gamma_1 \epsilon_1
 \biggl( \bm{Y}_{N_1} - \frac{Y^\eq_{N_1}}{Y^\eq_\chi} \bm{Y}_\chi \biggr).
 \\
 \bm{Y}_{N_2} &= \frac{45 \sqrt{10}}{2 \pi^5 g^{1/2}_* g^S_*}
 \frac{M_P m_{N_2} \tilde{\Gamma}_2}{m_{N_1}^3} x^3,
 \\
 \bm{Y}_\chi &= \frac{135}{2 \pi^6 g_* g_*^S}
 \frac{M_P^2 m_{N_2}^2 \tilde{\Gamma}_2 ( \Gamma_{N_2 \to N_1 \chi} + \Gamma_2)}{m_{N_1}^6} x^6,
 \\
 \bm{Y}_{N_1} &= \frac{45 \sqrt{10}}{2 \pi^5 g^{1/2}_* g^S_*}
 \frac{M_P \tilde{\Gamma}_1}{m_{N_1}^2} x^3
 + \frac{135}{2 \pi^6 g_* g_*^S}
 \frac{M_P^2 m_{N_2}^2 \tilde{\Gamma}_2 ( \Gamma_{N_2 \to N_1 \chi} + \Gamma_2)}{m_{N_1}^6} x^6.
\end{align}
$z_1$ is given by using these $\bm{Y}_{N_2}$ and $\bm{Y}_\chi$ as
\begin{align}
 z_1 = \mathop{\mathrm{min}} ( z_{N_2}, z_\chi),
 \quad
 \bm{Y}_{N_2} (z_{N_2}) = \frac{45}{\pi^4 g^S_*},
 \quad
 \bm{Y}_\chi(z_\chi) = \frac{45}{2\pi^4 g^S_*},
\end{align}
and $z_{N_2}$ ($z_{\chi}$) is regarded as the time of $N_2$ ($\chi$) being close to the thermal equilibrium.
The washout gives the exponential suppression through
$e^{ - \int dx' \, W_2(x')}$.
On the other hand,
if the $N_1$ process is dominant,
the asymmetry is mainly produced by the ordinary freeze-in mechanism via $N_1$ decay
such as the FIMP dark matter \cite{Hall:2009bx} (top-right panel in Fig.~\ref{fig:sol-BE}).
The relic of the lepton asymmetry is written as
\begin{align}
 Y^{\mathrm{FI}}_{\Delta L} (\infty) \approx
 \frac{135 \sqrt{10} }{2 \pi^5 g^{1/2}_* g_*^S}
 \frac{M_P \Gamma_1 \epsilon_1}{m_{N_1}^2}
 \int_0^\infty dx\, x^3 K_1 (x)
 =
 \frac{405 \sqrt{10}}{4 \pi^4 g^{1/2}_* g_*^S}
 \frac{M_P \Gamma_1 \epsilon_1}{m_{N_1}^2}.
 \label{eq:YFI(infty)}
\end{align}

If the washout is strong and dilutes the relic from the $N_1$ decay
(bottom right panel in Fig.~\ref{fig:sol-BE}),
the lepton asymmetry is evaluated as
\begin{align}
 \mathcal{Y}^{\mathrm{WO}}_{\Delta L }(\infty) = \int_0^\infty dx'\, \mathcal{F}_1(x')
 \exp \biggl[ - \int_{x'}^{\infty} dx'' \, W_1(x'') \biggr],
 \label{eq:mathcalY-WO}
\end{align}
where the washout function of this case is given by
\begin{align}
 W_1 \coloneqq \frac{1}{2 H x} \frac{K_1 (m_{N_2} x /m_{N_1})}{K_2 (m_{N_2} x / m_{N_1})} ( \tilde{\Gamma}_2 + \Gamma_2 ) \frac{Y^\eq_{N_2}}{Y^\eq_\LL}
 + \frac{1}{2 H x} \frac{K_1 (x)}{K_2(x)} ( \tilde{\Gamma}_1 + \Gamma_1) \frac{Y^\eq_{N_1}}{Y^\eq_\LL},
\end{align}
and we introduce the following function
\begin{align}
 \mathcal{F}_1\coloneqq&
 \frac{1}{H x} \frac{K_1(m_{N_2} x/ m_{N_1})}{K_2( m_{N_2} x /m_{N_1})}
 \Gamma_2 \epsilon_2 \biggl( \frac{ \tilde{\Gamma}_2 + \Gamma_2}{m_{N_2}} \biggr) Y^\eq_{N_2}
 + \frac{1}{H x} \frac{K_1(x)}{K_2(x)}
 \Gamma_1 \epsilon_1 \biggl( \frac{\tilde{\Gamma}_1 + \Gamma_1}{m_{N_1}} \biggr) Y^\eq_{N_1}.
\end{align}

\medskip

\subsection{Parameter space}

Based on the analysis of the Boltzmann equations in the previous
subsection, we discuss the parameter spaces of the model where the lepton
asymmetry is properly generated. The Majorana masses of the RH
neutrinos are fixed to 
$m_{N_1}=10^{10}$ GeV, 
$m_{N_2}=10^{12}$ GeV, and 
$m_{N_3}=10^{14}$ GeV\@.
$N_3$ is heavy so that it is assumed not to join the
initial and final states of our considering processes.
The scalar $\chi$ is light and its mass is assumed in this work to be 
$m_\chi = 10^{-3}$ GeV, but its detailed value does not largely affect
the asymmetry. 
As for the CI parameters (the degrees of freedom of neutrino Yukawa
couplings), we consider the following two typical patterns of parameters:
\begin{enumerate}[label=(\Roman*)]
\item
Real CI parameters
\begin{align}
    \omega_{12} = \frac{1}{4},
    \quad
    \omega_{23} = \frac{3}{7},
    \quad
    \omega_{13} = 0
    \label{eq:Real-CI}
\end{align}

\item
Pure imaginary CI parameters
\begin{align}
    \omega_{12} = \frac{i}{4},
    \quad
    \omega_{23} = \frac{3i}{7},
    \quad
    \omega_{13} = 0.
 \label{eq:Pure-imaginary-CI}
\end{align}

\end{enumerate}
With these parameter sets,
the asymmetry from the SM particle loops vanish, 
$\tilde\epsilon^{(LH)}_i = 0$, found from Eq.~\eqref{eq:tildeepsilonLH}.

\begin{figure}[t]
    \centering
    \includegraphics[width=0.38\textwidth]{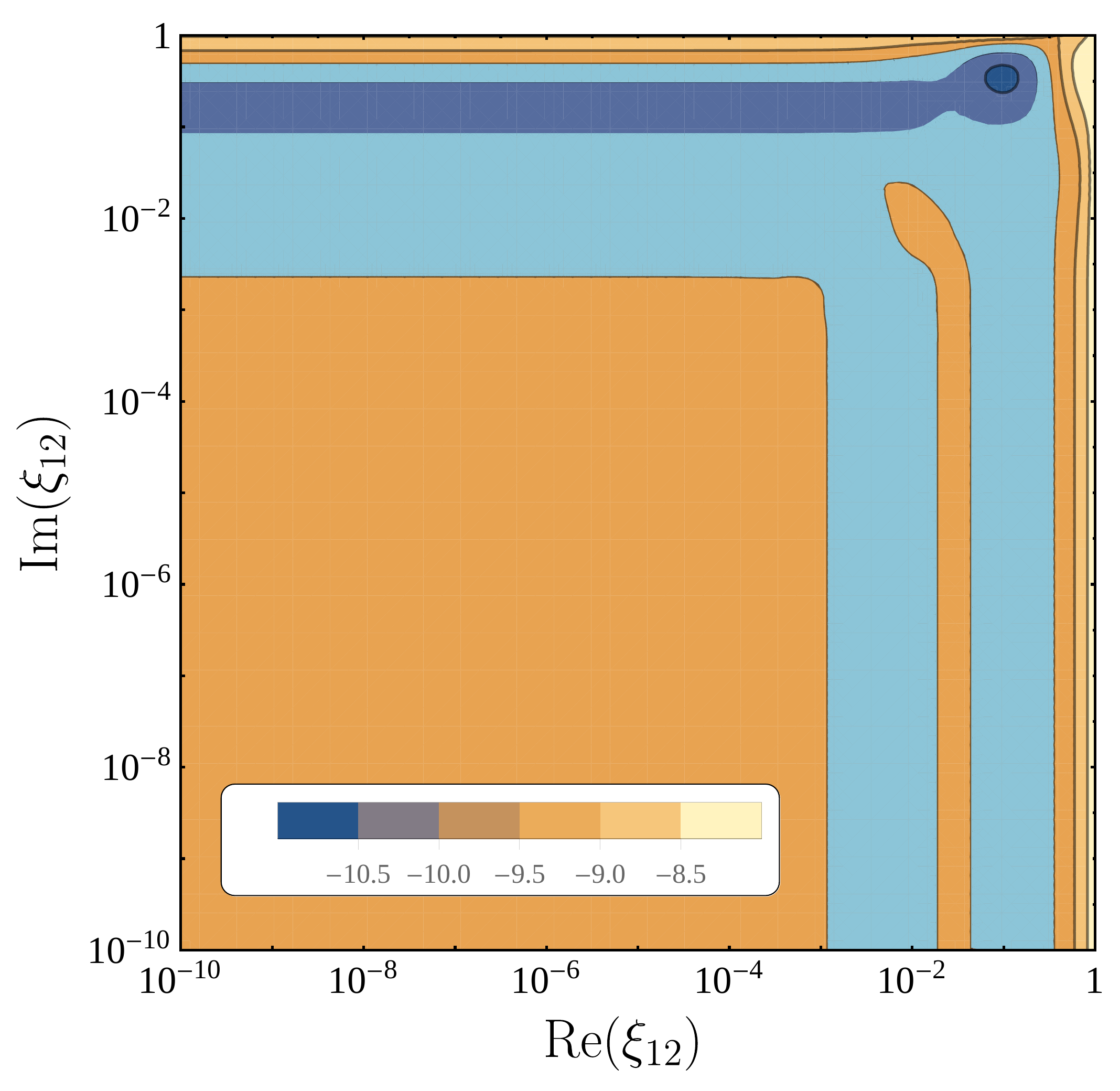}
    \quad 
    \includegraphics[width=0.38\textwidth]{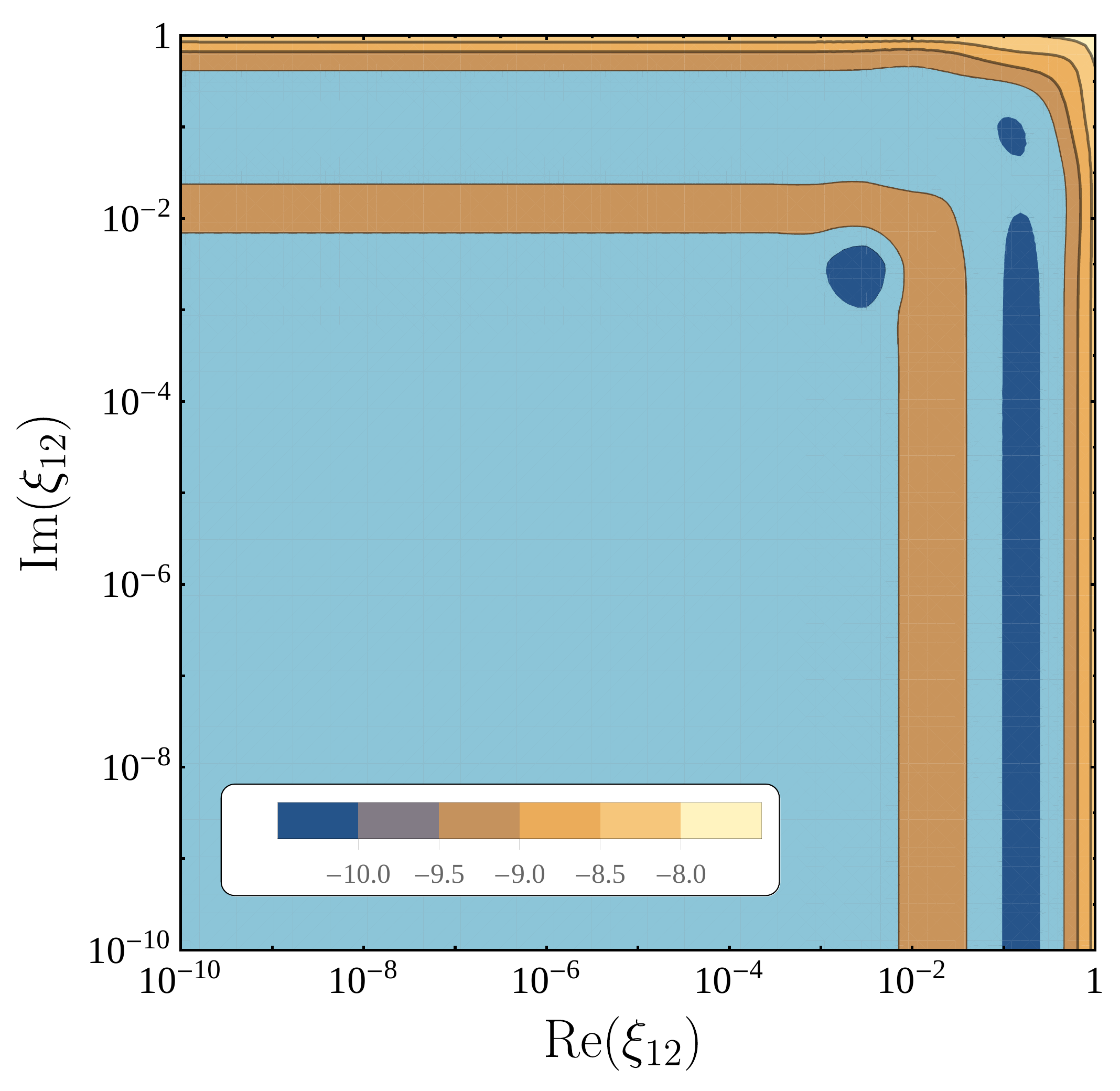}
    \\[2mm]
    \includegraphics[width=0.38\textwidth]{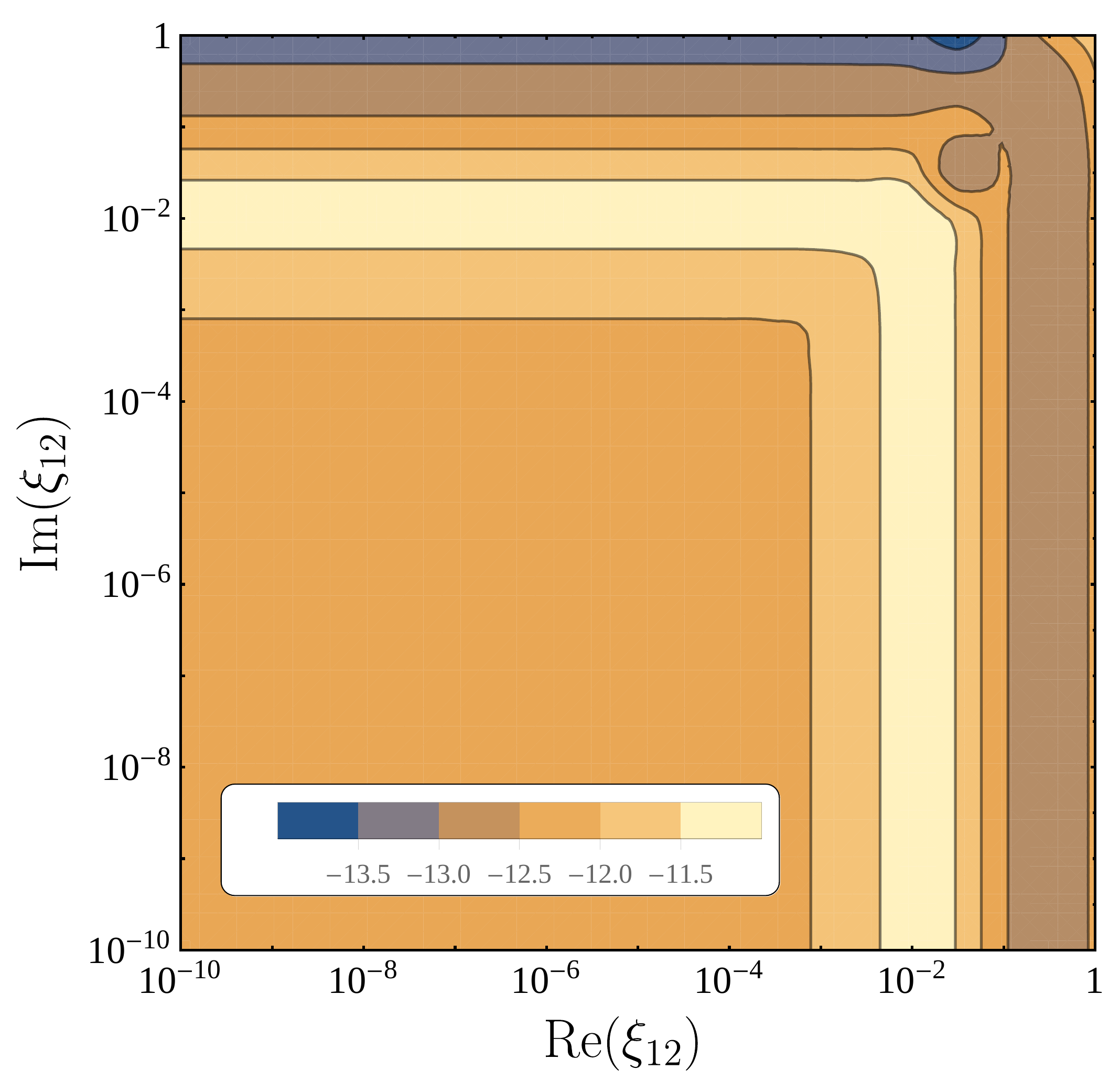}
    \quad
    \includegraphics[width=0.38\textwidth]{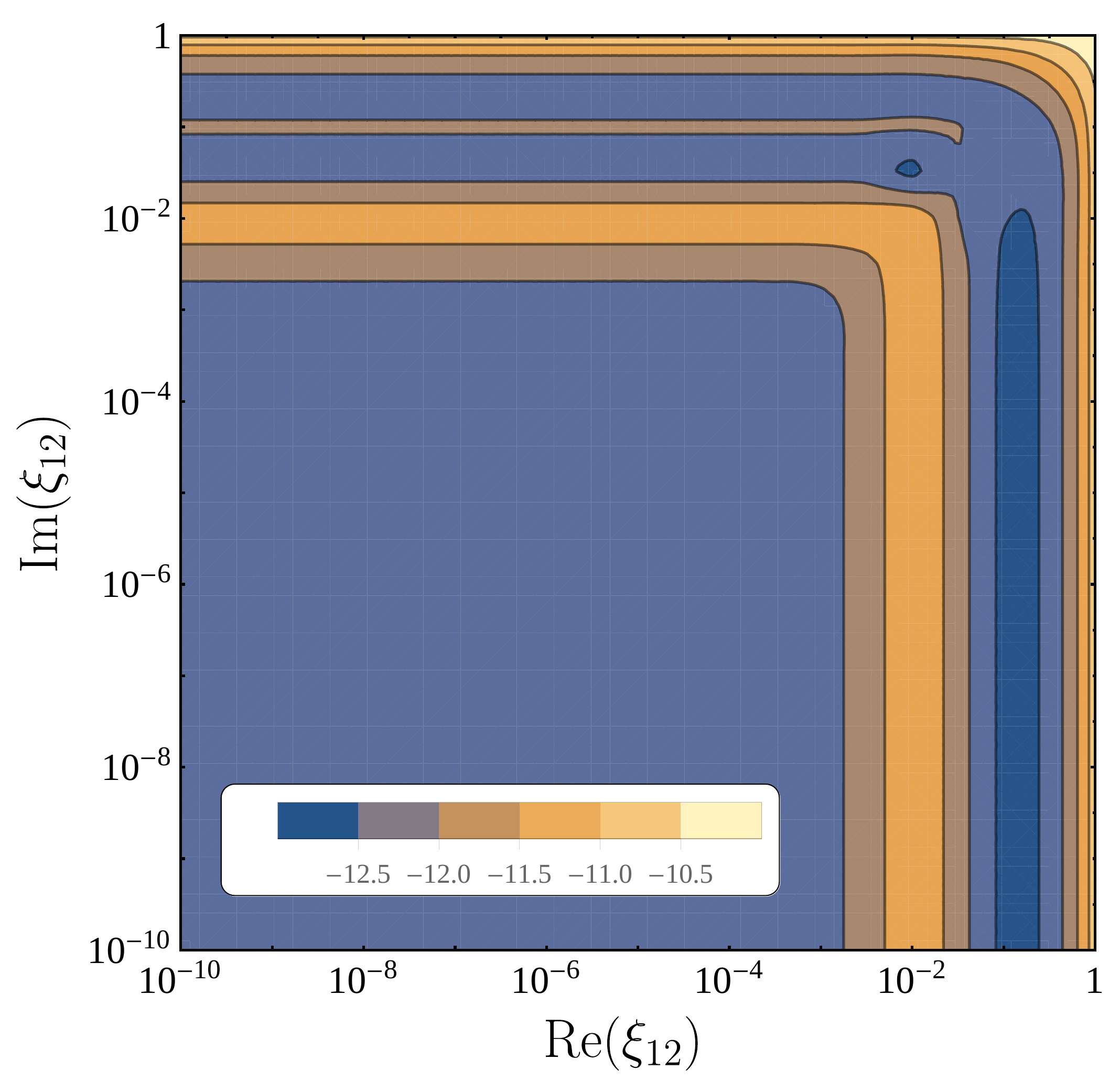}
    \caption{%
        The density plots of the lepton asymmetry in the $(\Re \xi_{12}, \Im \xi_{12})$ plane.
        The upper figures show the parameter space for the normal mass hierarchy and the real (imaginary) CI parameters in the left (right) figure.
        The similar figures are given for the inverted mass hierarchy in the lower panels.
        The light blue regions properly produce the observed baryon asymmetry in the Universe.
    }
\bigskip
\label{fig:Rxi12-Imxi12}
\end{figure}

Fig.~\ref{fig:Rxi12-Imxi12} shows the relic of the lepton asymmetry
$Y_{\Delta L}(\infty)$ for
\begin{align}
    \Re \xi = \begin{pmatrix}
        10^{-2} & \Re \xi_{12} \\
        \Re \xi_{12} & 10^{-1}
    \end{pmatrix},
    \qquad
    \Im \xi = \begin{pmatrix}
        10^{-1} & \Im \xi_{12} \\
        \Im \xi_{12} & 1 
    \end{pmatrix}.
\end{align}
The light blue regions properly produce the observed baryon
asymmetry. 
The CP asymmetry parameters Eqs.~\eqref{eq:N1eps11}, \eqref{eq:N1eps12} and \eqref{eq:N2eps11}--\eqref{eq:N2eps22}
are roughly determined by the large values in $\xi^2$,
which then lead to the contours seen in Fig.~\ref{fig:Rxi12-Imxi12}.

The left panel of Fig.~\ref{fig:Rxi12-Imxi12} is the parameter space
for the normal mass hierarchy of neutrinos and real CI parameters, 
where the washout effect is less dominant.
In this case, larger values of couplings lead larger lepton asymmetry 
through the freeze-in like production of $N_1$.
Then the lepton asymmetry reduces due to weaker couplings,
but the asymmetry becomes large again when $\xi_{12}$ is small enough.
This is because the freeze-in like production of $N_2$ is dominant.
If the washout effect works well as in the inverted mass hierarchy,
the lepton asymmetry is suppressed as shown in the middle and right panels
of Fig.~\ref{fig:Rxi12-Imxi12}. When the CI parameters are real,
the final value of the lepton asymmetry tends to converge to
a value slightly smaller than $10^{-10}$ if $\xi$ are feeble couplings,
which means the relic asymmetry is determined by the neutrino Yukawa couplings.
However the washout is relatively stronger if $\xi$ is larger, and the
relic is more suppressed. 
On the other hand,
the washout effect of neutrino Yukawa couplings works well and the
lepton asymmetry can only take a tiny value. 
In this case, an enough large coupling is needed for a large amount of
$Y_{\Delta L}$ produced in the early Universe
to explain the observed asymmetry even when it is washed out.
From these observations, the leptogenesis from the
three-body decay of RH neutrinos with a flavorful scalar does not
favor the inverted mass hierarchy of neutrinos.

\medskip

\begin{figure}[t]
    \centering
    \includegraphics[width=0.32\textwidth]{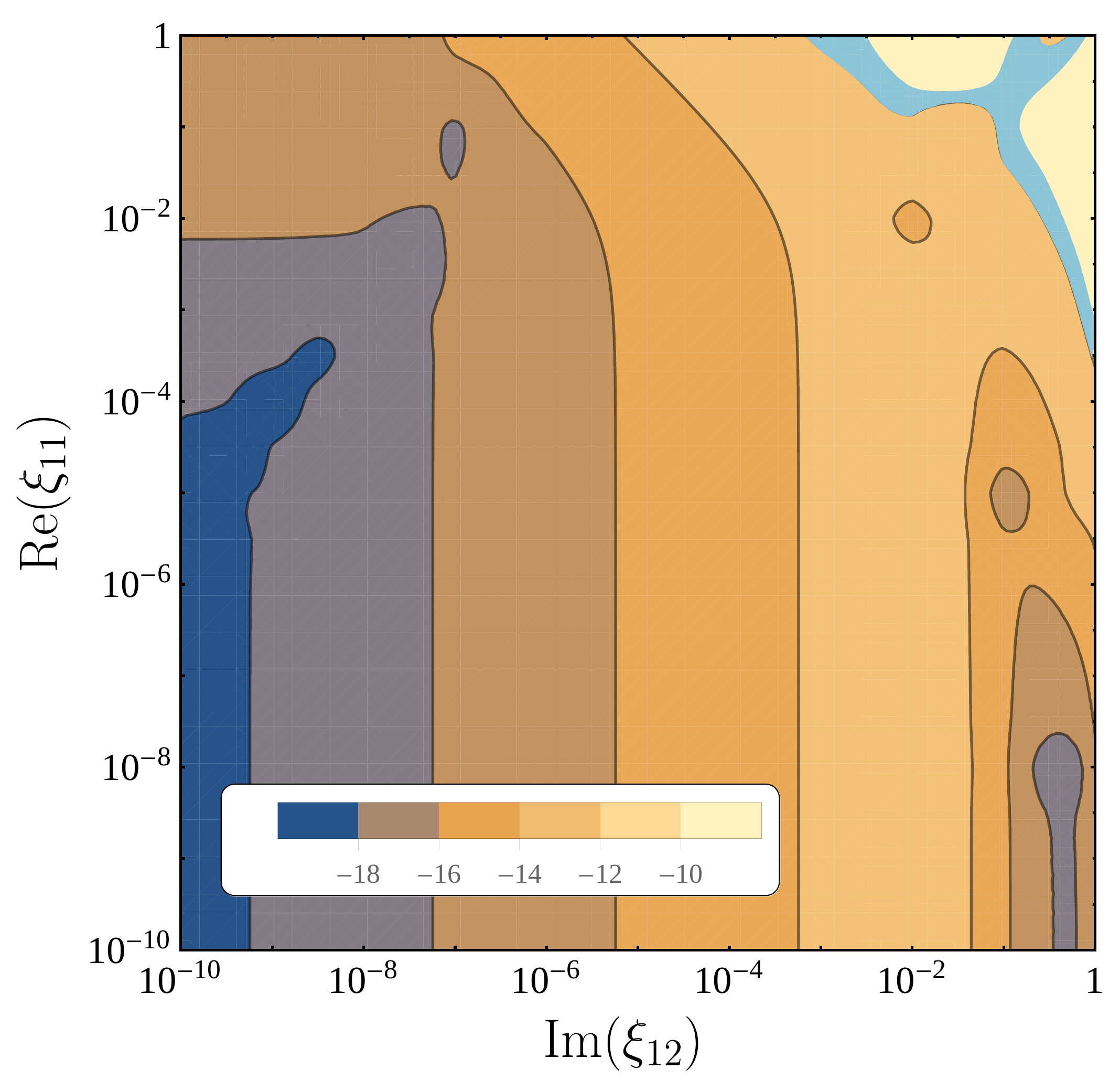}\ 
    \includegraphics[width=0.32\textwidth]{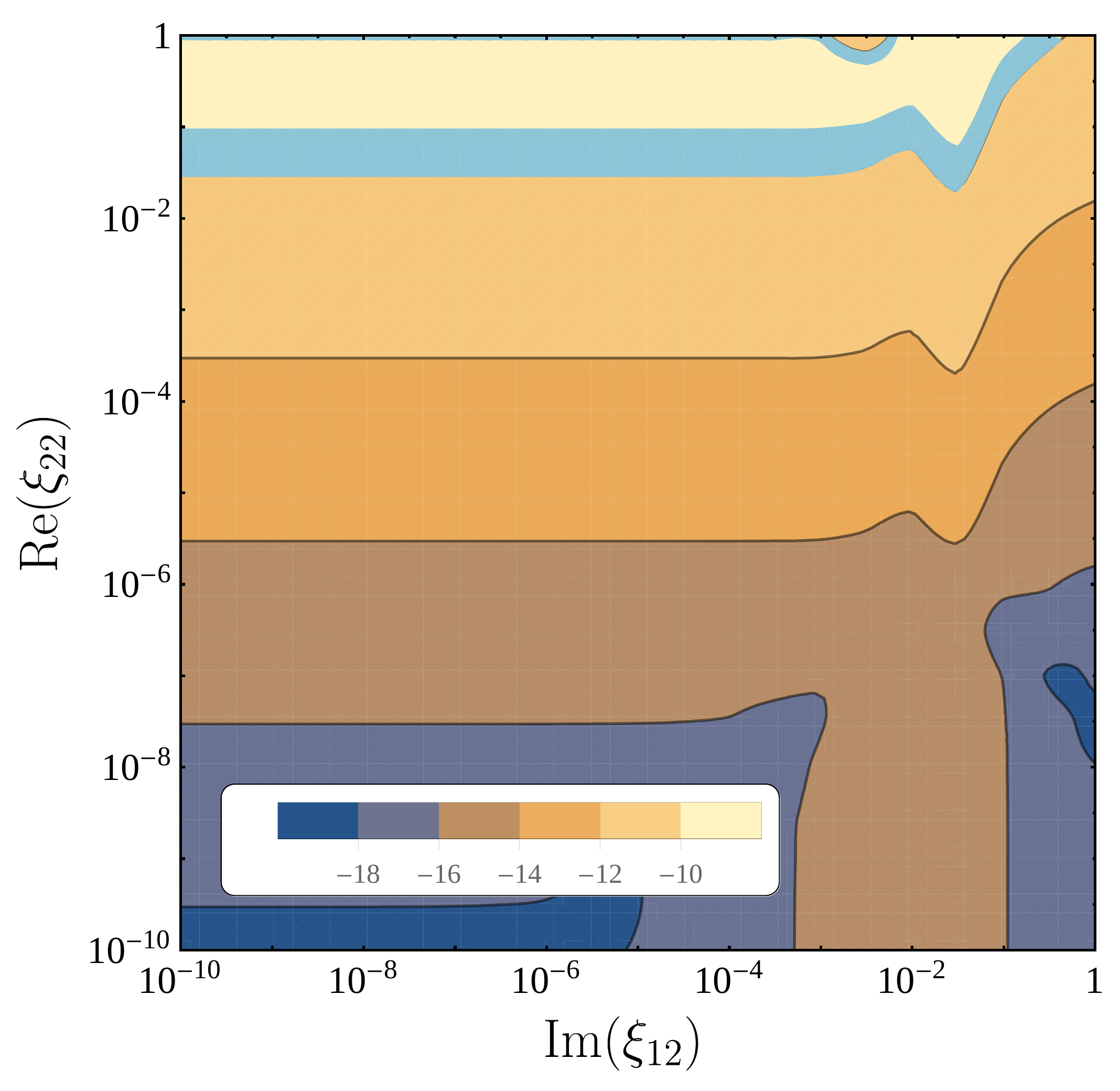}\ 
    \includegraphics[width=0.32\textwidth]{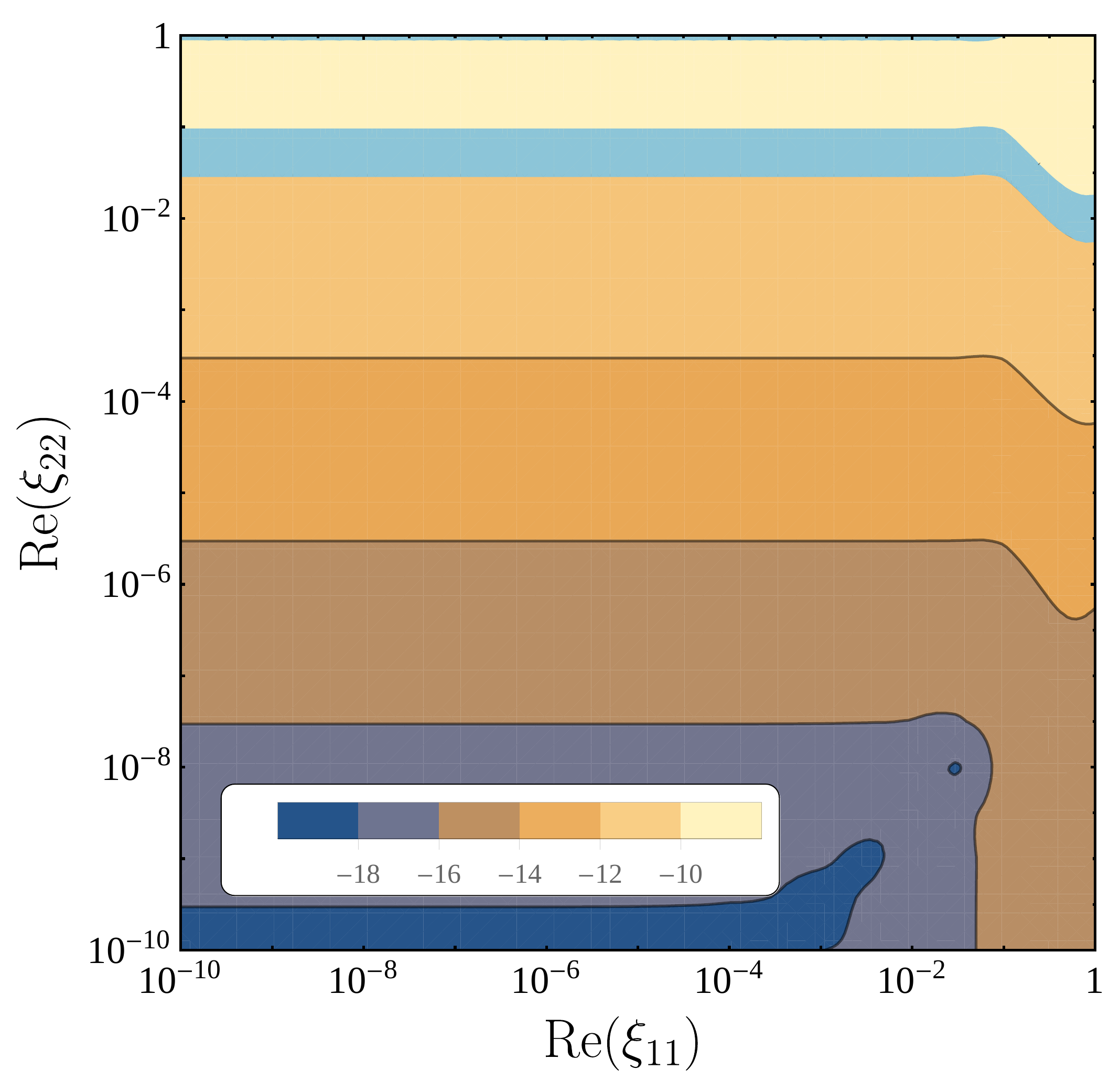}\ 
    \caption{%
        The density plots of the lepton asymmetry in the $(\Im \xi_{12}, \Re\xi_{11})$, $(\Im \xi_{12}, \Re \xi_{22})$ and $(\Re \xi_{11} ,\Re \xi_{22})$ planes with the parameterization~\eqref{eq:xi-pNGB+h-pert}. 
        The normal mass hierarchy and real CI parameters are adopted for realizing $Y_{\Delta L}(\infty) \sim 10^{-10}$.
    }
    \bigskip
    \label{fig:Imxi-Rexi-NH-CIre}
\end{figure}

We now discuss a singlet scalar $\chi$ and its general form of
couplings to the RH neutrinos. If $\chi$ is something like a
pseudo-Nambu-Goldstone boson
associated with the lepton number symmetry,
the imaginary parts of the diagonal couplings $\xi_{ii}$ tend to be
dominant. Keeping in mind this fact, we take the coupling $\xi$ as
\begin{align}
 \Re \xi = \begin{pmatrix}
 \Re \xi_{11} & 0 \\
 0 & \Re \xi_{22}
 \end{pmatrix},
 \qquad
 \Im \xi = \begin{pmatrix}
 10^{-2} & \Im \xi_{12} \\
 \Im \xi_{12} & 1
 \end{pmatrix}.
 \label{eq:xi-pNGB+h-pert}
\end{align}
The parameter spaces are shown in Fig.~\ref{fig:Imxi-Rexi-NH-CIre} when
$\Re \xi_{11}$, $\Re \xi_{22}$ and $\Im \xi_{12}$ are assumed to be
the modifications to the imaginary diagonal elements. 
In this situation, 
the lepton asymmetry is produced by the freeze-in mechanism of $N_1$
or $N_2$ as shown in the top panels of Fig.~\ref{fig:sol-BE}, 
and there exist the allowed regions between the parameter spaces of
over- and under-productions of the asymmetry.

\medskip

\begin{figure}[t]
    \centering
    \includegraphics[width=0.32\textwidth]{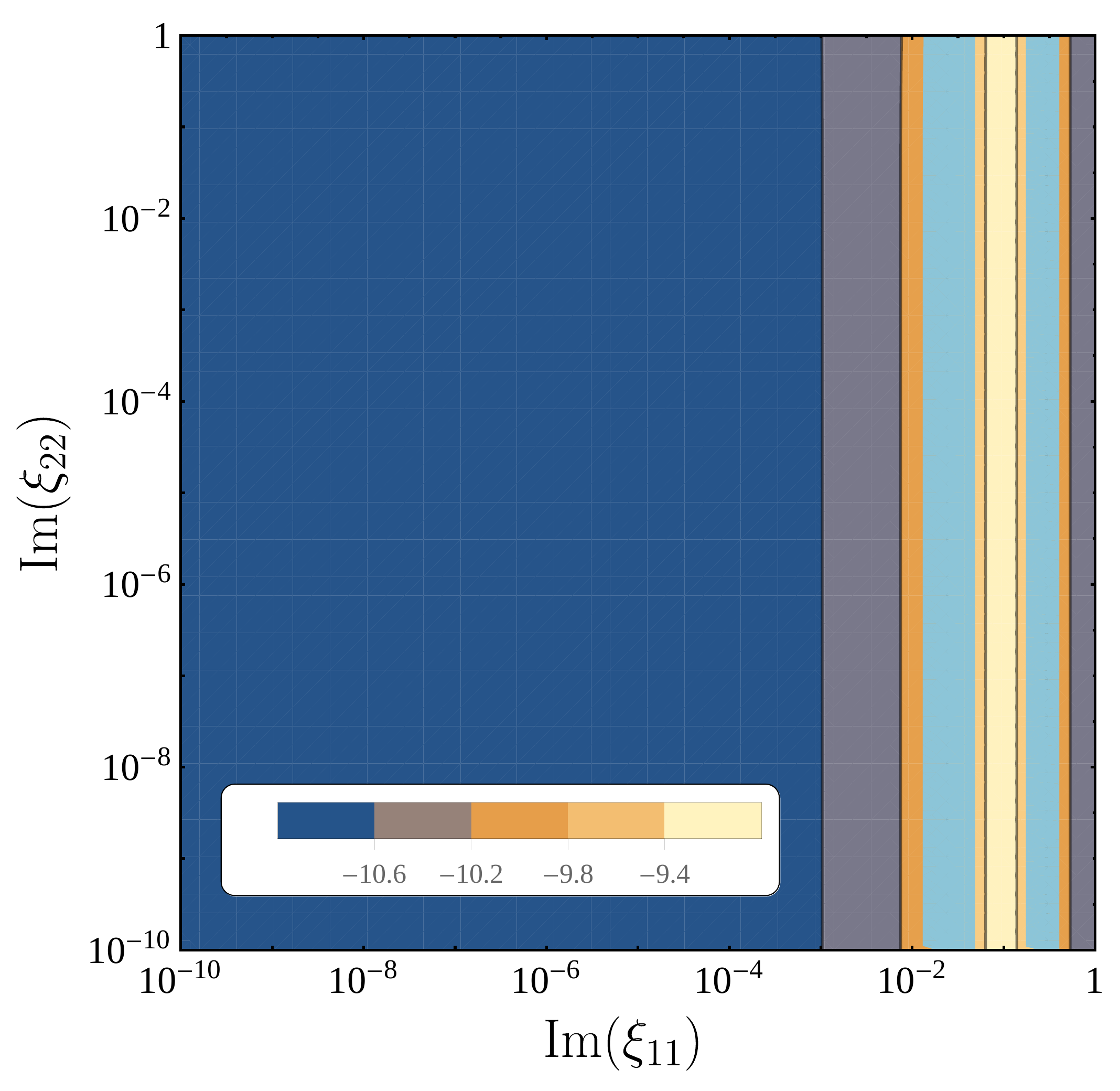}\ 
    \includegraphics[width=0.32\textwidth]{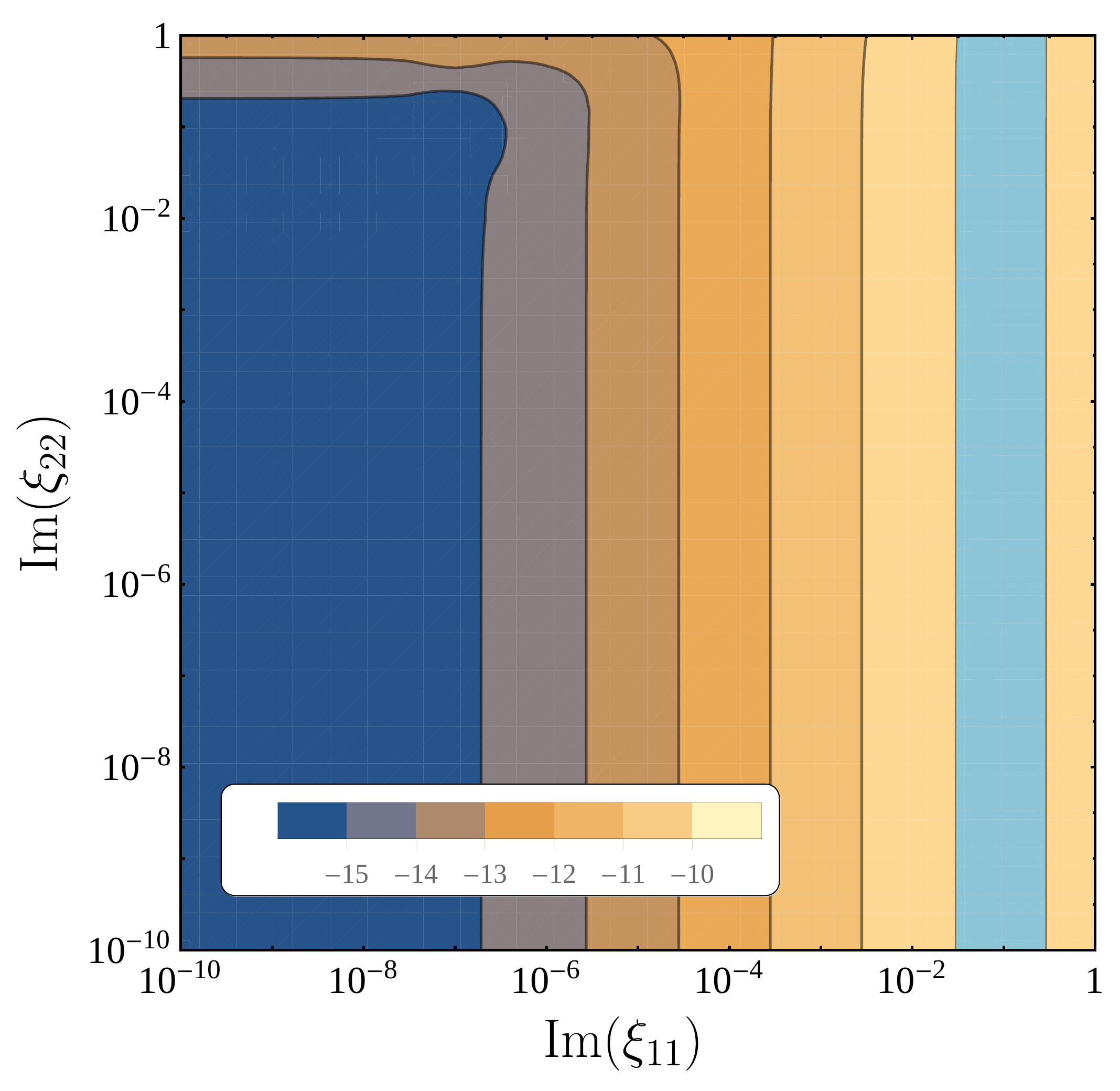}\ 
    \includegraphics[width=0.32\textwidth]{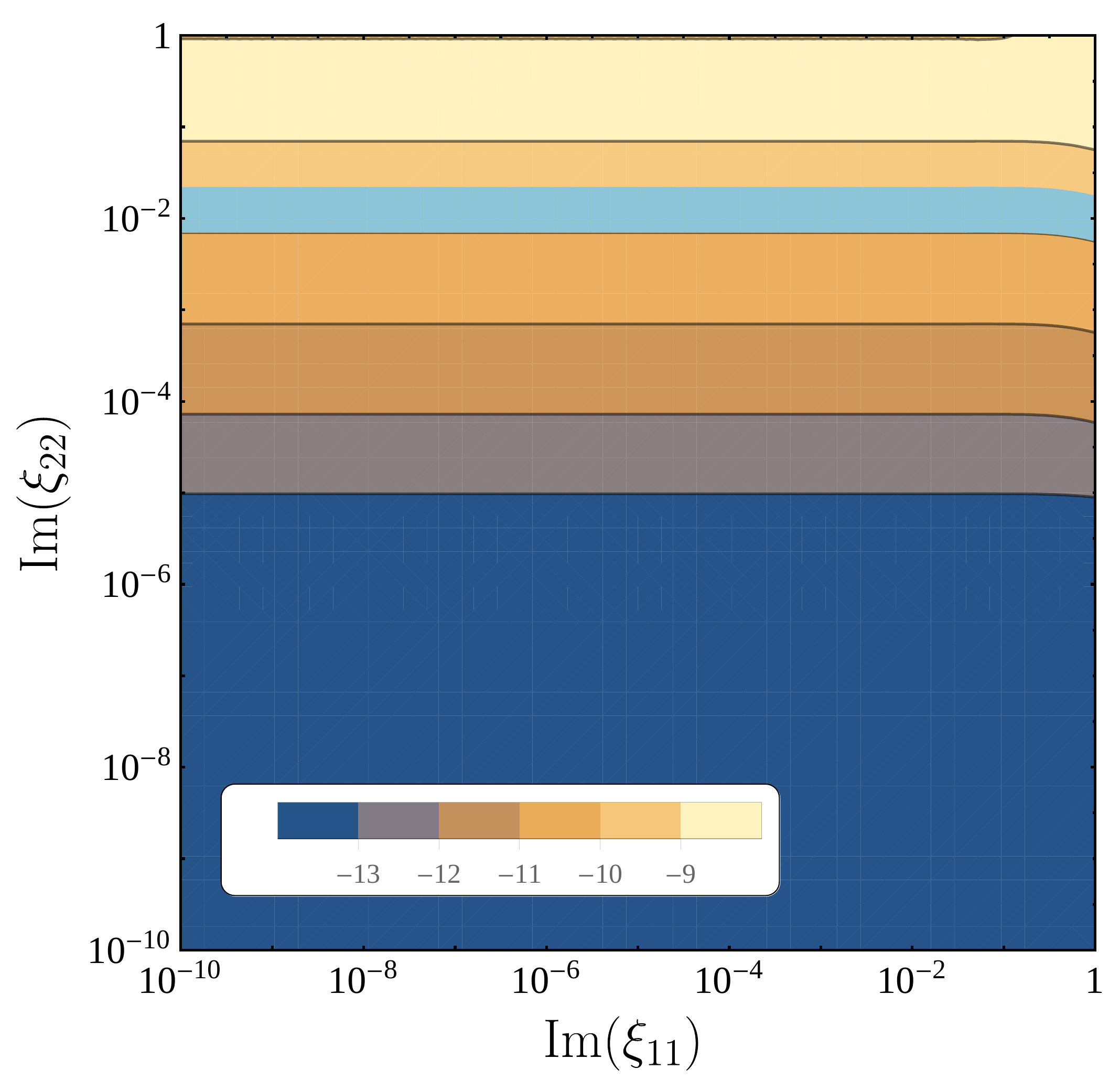}\ 
    \caption{%
        The density plots of the lepton asymmetry in the $(\Im \xi_{11}, \Im \xi_{22})$ plane for the normal mass hierarchy.
        (Left) $\Im \xi_{12} = 10^{-1/2}$, $\Re \xi_{12} =10^{-1/2}$.
        (Middle) $\Im \xi_{12} = 10^{-5}$, $\Re \xi_{12} = 10^{-1/2}$.
        (Right) $\Im \xi_{12} = 10^{-5}$, $\Re \xi_{22} =10^{-1/2}$.
    }
    \bigskip
    \label{fig:g11-g22}
\end{figure}

On the other hand, the parameter spaces of the imaginary diagonal
elements are shown in Fig.~\ref{fig:g11-g22}. 
For evaluating the lepton asymmetry in these planes,
$\Im \xi_{12}$ and one of $\Im \xi_{12}$ or $\Re \xi_{22}$ are fixed to
$10^{-1}$ or $10^{-5}$, and the normal mass hierarchy for the neutrino
masses is assumed. As seen from these figures, 
larger couplings lead to larger relic asymmetry, 
similar to the above cases of the normal mass hierarchy, 
and $\Im \xi_{11}$, $\Im\xi_{22}$ $ \sim 10^{-(1 \text{--} 2)}$ are favored.

\subsection{Property of $\chi$ scalar}
\label{sec:property-of-chi}

In the present work, the scalar field $\chi$ has two important
property that (i) it has the complex coupling $\xi_{ij}$ to the parent
decay particles and (ii) its remnant $Y_\chi$ in the present Universe
is large. That is a general result to have an appropriate order of
baryon asymmetry generated from the three-body decay including $\chi$.  

Let us first discuss about the coupling $\xi_{ij}$. The analysis in
the above shows that $\xi_{ij}$ should be complex-valued so that
the CP asymmetry is properly produced through the three-body decay.
Further the flavor-changing
components $\xi_{ij}$ ($i\neq j$) can play an important role for
the leptogenesis. These nature of couplings put some constraint on the
property of the field $\chi$, for example, a dynamical completion in
high-energy regime. 
\begin{itemize}
\item Nambu-Goldstone boson : \\
\phantom{xx} A simple example for scalar couplings to neutrinos is
the one whose vacuum expectation value (VEV) gives the masses of
RH neutrinos $N_i$. If such a scalar is complex, the coupling has
the lepton number symmetry (the chiral rotation of $N_i$) and the
scalar VEV breaks it. As a result of symmetry breaking, a
Nambu-Goldstone boson (NGB) appears
\cite{Chikashige:1980qk,Chikashige:1980ui,Gelmini:1980re}, which
couples to $N_i$ and can be identified to $\chi$. 
This is a typical and the simplest dynamical realization of the
present model. It is however noticed that, if the NGB 
nature is exact, $\chi$ is massless and its couplings to $N_i$ are
real and flavor diagonal, i.e., $\xi_{ij}= c_i\,\delta_{ij},~ c_i \in \mathbb{R}$. 
As mentioned above, such a too simple form of scalar
couplings is not suitable for generating the asymmetry.

\item Flavor-dependent interaction : \\
\phantom{xx} There are several ways to ameliorate the NGB problem, too
simple $\xi_{ij}$, found in the above simplest setup. The first is to
introduce multi scalars which couple to the RH neutrinos
$N_i$. These scalars generally couple to each other 
and develop nonzero VEVs. The resultant NGB, which is identified to
$\chi$, is a linear combination of (the phases of) these scalars. As a
result, the flavor structure becomes different between the masses and
the scalar couplings of RH neutrinos.\\
\phantom{xx} A more interesting realization is to assign
generation-dependent charges to leptons under some flavor
symmetry. The assignment is chosen so that 
the masses (and Yukawa couplings) are forbidden and then induced by
some VEVs of complex scalar, charged under the same flavor
symmetry~\cite{Froggatt:1978nt}. 
In this case, $\chi$ is possibly the NGB of flavor symmetry and its
coupling to leptons is determined by the flavor structure of lepton
masses and symmetry charges. Unless the masses and charges have
exactly the same structure, the $\chi$ couplings are generally flavor
dependent~\cite{Davidson:1981zd,Wilczek:1982rv} (and complex valued). Typical
examples may be constructed with flavor $U(1)$ symmetry for fermion mass
hierarchy, and non-universal anomaly-free lepton numbers such as $\mathbb{L}_\mu-\mathbb{L}_\tau$ and others~\cite{Bento:1991bc,Ma:2014qra}.\\
\phantom{xx} Along this line, we here present an example for obtaining general couplings $\xi_{ij}$. Let us consider a $U(1)$ symmetry under which the two-component RH neutrinos $\nu_{1,2}$ have the charges $q_{1,2}$ and a complex scalar $\Phi$ has the charge $-1$.
The symmetry-invariant operator for $\nu$'s is
\begin{align}
  \mathcal{L} \;=\; \sum_{i,j}
  \frac{1}{2}f_{ij}\Phi^{q_i+q_j}
  \nu_i^\rt \varepsilon\nu_j^{} +\text{h.c.}
\end{align}
The couplings $f_{ij}$ are generally complex-valued. With a
non-vanishing VEV, $\Phi=\frac{1}{\sqrt{2}}(v_\phi+\phi+i\chi)$, we have 
\begin{align}
  \mathcal{L} \;=\; \sum_{i,j}
  \frac{1}{2}\nu_i^\text{t}m_{ij}\varepsilon\nu_j^{}
  +\frac{1}{2}i\chi\,\nu_i^\rt h_{ij}\varepsilon\nu_j^{}
  +\text{h.c.} 
\end{align}
with $m_{ij}=f_{ij}(v_\phi/\sqrt{2})^{q_i+q_j}$ and $h_{ij}=(q_i+q_j)m_{ij}/v_\phi$.
 We have dropped the term of heavy $\phi$ field. The diagonalization $m_\text{diag}= U^\rt m U=\diag (m_{N_1},m_{N_2})$ leads
to the coupling $\eta=iU^\rt hU$, whose components are 
expressed by $m_{N_1}$, $m_{N_2}$, $m_{11}$ 
instead of the original couplings $f_{ij}$ as
\begin{align}
  \eta = \!\left(\begin{matrix}
  \frac{2m_{N_1}(q_1m_{N_2}-q_2m_{N_1}+m_{11}(q_2-q_1))}{v_\phi(m_{N_2}-m_{N_1})}
  & \!\!\!\! \frac{(q_2-q_1)(m_{N_1}+m_{N_2})(m_{N_2}-m_{11})^{1/2}
(m_{11}-m_{N_1})^{1/2}}{v_\phi(m_{N_2}-m_{N_1})} \\[1mm]
  \!\frac{(q_2-q_1)(m_{N_1}+m_{N_2})(m_{N_2}-m_{11})^{1/2}
(m_{11}-m_{N_1})^{1/2}}{v_\phi(m_{N_2}-m_{N_1})}
  & \frac{2m_{N_2}(q_2m_{N_2}-q_1m_{N_1}+m_{11}(q_1-q_2))}{v_\phi(m_{N_2}-m_{N_1})}
\end{matrix}\right)\!.
\end{align}
By rephasing $m_\text{diag}$ by $P$ to make the eigenvalues real, we find the coupling $\xi=P\eta P$ between the NGB $\chi$ and the four-component RH neutrinos denoted by $N_i$.
Due to generation-dependent charges, the mass $m_{ij}$ and the coupling $h_{ij}$ have different structure, which leads to the general complex form of $\xi$.

\item Radiative corrections : \\
\phantom{xx} The NGB property is not exactly hold if the original symmetry is
somehow violated via explicit breaking terms. A well-known example is
the scalar mass term which does not respect (global) symmetry, and
hence the NGB acuires its nonzero mass (the ``pion'' mass) at classical
level. With such explicit breaking, one generally expects to have 
non-vanishing radiative corrections to unusual NGB couplings and resolve
the NGB problem mentioned above. 
There are phenomenological analysis for large breaking, e.g., a heavy
NGB of the lepton number symmetry~\cite{Abe:2020dut,Matsumoto:2010hz},
where suitably size of corrections might be obtained.

\end{itemize}

\medskip

The large remnant of $\chi$ is another characteristic result of the
model. We here discuss two approaches to this issue: (i) additional
scalar interactions. (ii) a very light $\chi$. The first resolution
is to introduce additional interaction such that it suppresses
the scalar abundance somewhere in the thermal history, typically later than the leptogenesis epoch. However, 
when $\chi$ is the NGB of high-scale symmetry, its large decay constant
generally suppresses the $\chi$ couplings to the SM fields and cannot
give an enough suppression to the $\chi$ abundance. 
For example, when $\chi$ is the NGB of lepton number symmetry which couples to
the RH neutrinos, a typical ratio 
$\Gamma/ H (T=m_\chi)$ is $10^{-20} \, \xi^2 (m_\chi/\text{GeV})$. 
That may therefore lead to a situation where the scalar including
three-body decay is not a NGB-like pseudo-scalar $\chi$ but some heavy
real scalar $\rho$ (heavier than the electroweak scale). 
Dynamical examples
of $\rho$ are a partner of NGB (a radial fluctuation around a VEV),
a massive scalar in the multi scalar scenario, and so on. A main
difference between $\chi$ and $\rho$ is the interaction to other (SM)
fields. In particular, for the Higgs portal interaction, $\rho$
interacts with the portal quartic coupling $\lambda$, but $\chi$ has a
suppressed amplitude given by $\lambda E^2/(\text{VEV})^2$ where $E$
is the energy scale considered. Upon the decoupling $E\sim m_\rho$ or 
$m_\chi$, the latter interaction (for $\chi$) is too tiny to reduce
the abundance, and the former one (for $\rho$) can be used for
the suppression. Whether such a heavy scalar can be the dark matter component in the Universe and does not disturb successful leptogenesis depends on the model parameters and needs a further detailed analysis.

The second option is to assume that $\chi$ is very light and does not 
contribute much to the mass density of the present Universe. 
The relic abundance of $\chi$ is given by 
$\Omega_\chi h^2 = m_\chi s_0 Y_{\chi,0}/(\varepsilon_{c,0}/h^2)$
with the today entropy and critical energy densities,
$s_0=2891~\mathrm{cm}^{-3}$ and 
$\varepsilon_{c,0} = 5.16(h/0.7)^2$ $\mathrm{GeV}/\mathrm{m}^3$.
If this abundance is required to be less than the observed dark matter
density times a small ratio $\delta$, that implies the upper bound on
$m_\chi$, 
\begin{align}
    m_\chi < \delta\,\frac{\Omega_\text{DM}
    \varepsilon_{c,0}}{s_0 Y_\chi^\text{eq}} 
    = 1.89\times 10^{-7}\,\delta\; [\text{GeV}]
\end{align}
where we have assumed $Y_{\chi,0}$ is equal to the equilibrium value
$Y_\chi^\eq$.
For example, $\chi$ has a sub-eV mass for $\delta=10^{-3}$ but weighs
more than the current temperature. Furthermore the interaction of $\chi$
to the SM sector is generally suppressed by $\mathcal{O}(m_\nu)$. 
Such a very light and feebly-interacting scalar may be harmlessly
floating in the present Universe.

\bigskip

\section{Summary}
\label{sec:summary}

We have studied the lepton asymmetry produced by the RH neutrino
three-body decays with a singlet scalar field.
In order to cover the general case,
we considered the scalar and pseudo-scalar couplings between the scalar field and RH neutrinos.
We evaluated the RH neutrino two-body and three-body decay widths with the scalar field and the asymmetry parameters.
For the decay widths,
the $k = l =1$ contribution tends to be dominant in the lightest $N_1$ decay,
and the contribution of the resonant $k=l=1$ or the non-resonant $k= l =2$ is dominant depending on $\xi$ in the heavier $N_2$ decay.
For the asymmetry parameters,
not only usual cross terms but also the $k=l=1$ can be dominant to $\epsilon_1$,
and the $k=l=2$ term is the leading contribution to $\epsilon_2$.
In the latter two contributions,
the CP asymmetry comes from a single decay process,
which is characteristic of the existence of the flavorful scalar.
We derived the Boltzmann equations and discussed there are four typical patterns of the lepton asymmetry production
depending on the $N_1$ or $N_2$ decay process being dominant to the production and the washout effect being strong or weak.
Based on these analyses,
we show the parameter space of the model where the lepton asymmetry is properly generated.
For example, we find the leptogenesis
via the RH neutrino three-body decay in Fig.~\ref{fig:three-body-decays}
with the flavorful scalar does not favor the inverted mass
hierarchy due to the strong washout suppression, without a help of the
two-body decay asymmetry.

In this paper, we consider a simplified model with a single flavorful
scalar coupling to the RH neutrinos and investigate the possibility
of the leptogenesis via the three-body decay with specific values of couplings.
Pursuing the ultraviolet origin of this additional scalar such as a
pseudo-Nambu-Goldstone boson and a more detailed analysis of flavor
dynamics are important and left for future study.

\section*{Acknowledgments}
\noindent
The authors thank to Takashi Toma and Takahiro Yoshida for useful comments and discussions.
This work is supported by JSPS Grant-in-Aid for Scientific Research KAKENHI Grant No.\ JP18H01214, JP20K03949, JP20J11901, and 
JSPS Overseas Research Fellowships (YA).

\bigskip

\appendix

\section{Three-body decay widths}
\label{sec:3-width}

We consider for simplicity the two-generation case $N_{1,2}$. 
The generalization to more generations is straightforward.
We assume $m_\chi\ll m_{N_i}$ and no large hierarchy among $m_{N_i}$. 
The masses of charged leptons and Higgs boson are dropped in the following formulae. 
The three-body decay width of the lightest RH neturino $N_1$ is found from the full form \eqref{eq:Gamma_N_itoLHchi} after the phase space integral as
\begin{align}
  \Gamma_{N_1\to L_jH\chi} = \sum_{k,\,l} y^\nu_{lj} y^{\nu *}_{kj} \Big(
  \xi_{1l}^*\xi_{1k} G_{kl}^A +
  \xi_{1l}\xi_{1k}^* G_{kl}^B +
  \xi_{1l}^*\xi_{1k}^* G_{kl}^C +
  \xi_{1l}\xi_{1k} G_{kl}^D  \Big) \,=\, \sum_{k,\,l} \Gamma_1^{(kl)},
  \label{eq:gammaN1} 
\end{align}
where $k,\,l$ indicate the contributions to the amplitudes which
contain the intermediate states are $N_{k,l}$, respectively. For 
example, $k\neq l$ means the cross term in the amplitude
squared. Each piece $G^{A,B,C,D}$ reads from the full form
\eqref{eq:Gamma_N_itoLHchi} and not explicitly given here. 
We instead show the exact form of
$\Gamma_1^{(kl)}$, corresponding $\mathcal{M}_k^*\mathcal{M}_l$, where
$\mathcal{M}_m$ is the $N_1$ three-body decay amplitude with the
intermediate state $N_m$. The result is found in the limit of light
$m_\chi$ as 
\begin{align}
	\Gamma_1^{(11)} =&~
  	\frac{1}{128\pi^3} y^\nu_{1j} y^{\nu *}_{1j}
  	\big[\,|\xi_{11}|^2m_{N_1} +\Re (\xi^2_{11}\tilde{m}'_{N_1})\big]
 	\ln \Bigl( \frac{m_{N_1}}{m_\chi} \Bigr), 
    \\
  	\Gamma_1^{(12)} =&~ \frac{1}{512\pi^3 m_{N_1}^3} 
   	y^\nu_{2j} y^{\nu *}_{1j} \bigg[ 
   	\xi_{11} \xi_{12}^* \tilde{m}_{N_1} \tilde{m}_{N_2}^*
   	\Big[\Big( m_{N_1}^2+m_{N_2}^2\Big)
   	\ln \Big(\frac{m_{N_2}^2}{m_{N_2}^2-m_{N_1}^2}\Big) 
   	-m_{N_1}^2 \Big]
    \nn \\
   	& \quad + 2m_{N_1} (\xi_{11} \xi_{12} \tilde{m}_{N_1} +
   	\xi_{11}^* \xi_{12}^* \tilde{m}_{N_2}^*)
   	\Big[ m_{N_2}^2 
   	\ln \Big(\frac{m_{N_2}^2}{m_{N_2}^2-m_{N_1}^2}\Big)
   	-m_{N_1}^2 \Big]
    \nn \\
   	& \quad +\xi_{11}^*\xi_{12} \Big[ 
   	m_{N_2}^2 (m_{N_1}^2+m_{N_2}^2) 
   	\ln \Big(\frac{m_{N_2}^2}{m_{N_2}^2-m_{N_1}^2}\Big) 
   	-m_{N_1}^2 m_{N_2}^2 -\frac{3}{2} m_{N_1}^4 \Big] \,\bigg] ,
    \\
  	\Gamma_1^{(21)} =&~ \Gamma_1^{(12)*} ,
    \\ 
  	\Gamma_1^{(22)} =&~ \frac{1}{512\pi^3 m_{N_1}^3}
  	y^\nu_{2j} y^{\nu *}_{2j} \bigg[ 
  	|\xi_{12}|^2 \Big[ (5 m_{N_2}^4-m_{N_1}^4) 
   	\ln \Big(\frac{m_{N_2}^2}{m_{N_2}^2-m_{N_1}^2}\Big) 
   	- 5m_{N_1}^2 m_{N_2}^2 -\frac{5}{2}m_{N_1}^4 \Big]  \nn \\
   	& \quad + 4 \Re(\xi_{12}^2 \tilde{m}_{N_2}) m_{N_1}  
   \Big[ (2 m_{N_2}^2-m_{N_1}^2) 
   \ln \Big(\frac{m_{N_2}^2}{m_{N_2}^2-m_{N_1}^2}\Big) 
   -2 m_{N_1}^2  \Big] \, \bigg] .
\end{align}
The quantities $\tilde{m}_N$ and $\tilde{m}'_N$ are 
$\bar{m}_N$ in the $N$ propagator evaluated at the pole and its imaginary replaced with $\gamma_N$, respectively.
When $N_2$ is much heavier than $N_1$, the expressions are reduced to
\begin{align}
    \Gamma_1^{(12)} = \Gamma_1^{(21)*}  \approx&~ 
    \frac{m_{N_1}\tilde{m}_{N_2}^*}{1024\pi^3 m_{N_2}^2} 
    y^\nu_{2j} y^{\nu *}_{1j} \,\xi_{12}^* 
    (3 \xi_{11}\tilde{m}_{N_1}+2\xi_{11}^* m_{N_1}),
    \\
    \Gamma_1^{(22)} \approx&~ 
    \frac{m_{N_1}^3}{768\pi^3 m_{N_2}^2} 
    y^\nu_{2j} y^{\nu *}_{2j} |\xi_{12}|^2.
\end{align}
The last equation implies that the $k=l=2$ mode does not induce CP
asymmetry at the leading order and then highly suppressed for the
$N_1$ decay.

For a heavier RH neutrino, its three-body decay can meet the
resonance around the mass of a lighter intermediate state, and 
then the width is largely enhanced. In the present case, the $N_2$
decay width $\Gamma_{N_2\to L_jH\chi}$ has the enhancement both for
$k=l=1$ and the sum of cross terms, namely, $\Gamma_2^{(11)}$ and
$\Gamma_2^{(12)}+\Gamma_2^{(21)}$ in the similar 
notation as \eqref{eq:gammaN1}. These on-shell contributions to the
decay width are evaluated from the full form
\eqref{eq:Gamma_N_itoLHchi} with the narrow width approximation. For
the $k=l=1$ part, we obtain 
\begin{align}
  \Gamma_2^{(11)} =
  \frac{m_{N_1}}{256\pi^2}
  \frac{m_{N_2}^2-m_{N_1}^2}{m_{N_2}^3\Gamma_{N_1}}\,
  y^\nu_{1j} y^{\nu *}_{1j} \Big[
  |\xi_{12}|^2 (m_{N_1}^2+m_{N_2}^2)
  +2 \Re(\xi_{12}^2\tilde{m}_{N_1})m_{N_2} \Big] ,
\end{align}
where $\Gamma_{N_1}$ in the denominator indicates $N_1$ is the real
intermediate state. Compared with the decay width for $N_2\to N_1\chi$
given in \eqref{eq:N2N1chi}, the on-shell $k=l=1$ contribution is found
\begin{align}
  \Gamma_2^{(11)} \approx \frac{m_{N_1}}{16\pi\Gamma_{N_1}}
  y^\nu_{1j} y^{\nu *}_{1j}\, \Gamma_{N_2\to N_1\chi} \,\approx\,
  \frac{1}{2}\Gamma_{N_2\to N_1\chi} .
\end{align}
For the cross-term part, a similar evaluation leads to the on-shell
resonant contribution,
\begin{align}
  \Gamma_2^{(12)}+\Gamma_2^{(21)} =&~ 
  \frac{m_{N_1}}{256\pi^2 m_{N_2}^3(m_{N_2}^2-m_{N_1}^2)} \bigg[
  \Gamma_{N_2} \Big[
  \tfrac{1}{2}\Re(y^\nu_{2j} y^{\nu *}_{1j}\xi_{22}^*\xi_{12})
  (m_{N_1}^2+m_{N_2}^2)^2   \nn \\
  & \quad
  -4\Im(y^\nu_{2j} y^{\nu *}_{1j}\xi_{12}^*)\Im(\xi_{22})
  m_{N_1}m_{N_2}(m_{N_1}^2+m_{N_2}^2) 
  +2\Re(y^\nu_{2j} y^{\nu *}_{1j}\xi_{22}\xi_{12})
  m_{N_1}^2m_{N_2}^2  \Big]  \nn  \\
  & \quad
  -\big[ \Im(y^\nu_{2j} y^{\nu *}_{1j}\xi_{22}\xi_{12}^*)m_{N_1}
  +\Im(y^\nu_{2j} y^{\nu *}_{1j}\xi_{22}^*\xi_{12}) m_{N_2} \big]
  (m_{N_2}^4-m_{N_1}^4)   \nn \\
  & \quad
  -2\big[\Im(y^\nu_{2j} y^{\nu *}_{1j}\xi_{22}\xi_{12})
  m_{N_1} +\Im(y^\nu_{2j} y^{\nu *}_{1j}\xi_{22}^*\xi_{12}^*)
  m_{N_2}\big]m_{N_1}m_{N_2}(m_{N_2}^2-m_{N_1}^2)  \,\bigg].
\end{align}
Finally the non-resonant part, $k=l=2$, gives 
\begin{align}
  \Gamma_2^{(22)} =
  \frac{1}{128\pi^3} y^\nu_{2j} y^{\nu *}_{2j}
  \big[ \,|\xi_{22}|^2 m_{N_2} + \Re (\xi^2_{22}\tilde{m}'_{N_2})\big]
  \ln \Bigl( \frac{m_{N_2}}{m_\chi} \Bigr) .
\end{align}

\bigskip

\section{CP asymmetry (width differences)}
\label{sec:cpasym}

For a real particle (a real scalar, a Majorana fermion, etc), the
decay to anti-particle final states is obtained by replacing all
coupling constants with their complex conjugates in the corresponding decay
amplitude to particle final states. In the present case, the
three-body decay of $N_i$ to the anti-lepton $\bar{L}_j$, the conjugate of
Higgs boson $\bar{H}$, and the real scalar $\chi$ is described by
$y^\nu\leftrightarrow y^{\nu *}$ and $\xi\leftrightarrow\xi^*$
(and $P_R\leftrightarrow P_L$) everywhere in the amplitude.
The CP asymmetry is induced at the decay, proportionally to the
difference of decay widths to particles and corresponding
anti-particles. As a result, the asymmetry is originated from the
pieces in the decay widths which are not real with respect to coupling
constants.

The decay width to anti-particles is defined in a similar way to the
decay to particles discussed in the previous section, namely,
\begin{align}
  \Gamma_{N_1\to L_j^c H^\dagger\chi} = 
  \sum_{k,\,l} \bar{\Gamma}_1^{(kl)} .
\end{align}
The differences of partial widths are written as 
$\Delta\Gamma_1^{(kl)}=\Gamma_1^{(kl)}-\bar\Gamma_1^{(kl)}$. From the
explicit forms for $\Gamma_1^{(kl)}$ and the replacement rule mentioned
above, we obtain
\begin{align}
    \Delta\Gamma_1^{(11)} =&~ 
    \frac{\gamma_{N_1}}{128\pi^3} y^\nu_{1j} y^{\nu *}_{1j}
    \Im (\xi_{11}^2) \ln \Bigl( \frac{m_{N_1}}{m_\chi} \Bigr),   \\
    \Delta\Gamma_1^{(12)}+\Delta\Gamma_1^{(21)} =&~ 
    \frac{1}{256\pi^3 m_{N_1}^3} \bigg[
    \Im(y^\nu_{2j} y^{\nu *}_{1j}\xi_{11} \xi_{12}^*)
    (\Gamma_{N_1}m_{N_2}-\Gamma_{N_2}m_{N_1})  \nn \\
    &\qquad  \times \Big[\Big( m_{N_1}^2+m_{N_2}^2\Big) 
    \ln \Big(\frac{m_{N_2}^2}{m_{N_2}^2-m_{N_1}^2}\Big) -m_{N_1}^2 \Big]  \nn \\
    & \quad + m_{N_1} \big[
    \Im(y^\nu_{2j} y^{\nu *}_{1j}\xi_{11} \xi_{12})\Gamma_{N_1} - 
    \Im(y^\nu_{2j} y^{\nu *}_{1j}\xi_{11}^* \xi_{12}^*) \Gamma_{N_2} \big]
    \nn \\
    &\qquad \times \Big[ m_{N_2}^2 
    \ln \Big(\frac{m_{N_2}^2}{m_{N_2}^2-m_{N_1}^2}\Big)
    -m_{N_1}^2 \Big] \,\bigg] ,  \\
    \Delta\Gamma_1^{(22)} =&~  
    \frac{\Gamma_{N_2}}{128\pi^3 m_{N_1}^2}
    y^\nu_{2j} y^{\nu *}_{2j}
    \Im(\xi_{12}^2)  
    \Big[ (2 m_{N_2}^2-m_{N_1}^2) 
    \ln \Big(\frac{m_{N_2}^2}{m_{N_2}^2-m_{N_1}^2}\Big) 
    -2 m_{N_1}^2  \Big]  .
\end{align}
When $N_2$ is much heavier than $N_1$, the expressions are reduced to
\begin{align}
  \Delta\Gamma_1^{(12)}+\Delta\Gamma_1^{(21)} \approx&~ 
  \frac{-m_{N_1}^2\Gamma_{N_2}}{512\pi^3 m_{N_2}^2} 
  \Im\big[y^\nu_{2j} y^{\nu *}_{1j}\xi_{12}^*(3\xi_{11} +2\xi_{11}^*)
  \big] ,  \\
  \Delta\Gamma_1^{(22)} \approx&~  
  \frac{m_{N_1}^4\Gamma_{N_2}}{768\pi^3 m_{N_2}^4}
   y^\nu_{2j} y^{\nu *}_{2j}
   \Im(\xi_{12}^2)  .
\end{align}

In a similar way, we have the width differences for the $N_2$ resonant
and non-resonant three-body decay,
\begin{align}
    \Delta\Gamma_2^{(11)} =&~ 
    \frac{m_{N_1}}{128\pi^2} \frac{m_{N_2}^2-m_{N_1}^2}{m_{N_2}^2}\,
    y^\nu_{1j} y^{\nu *}_{1j} \Im(\xi_{12}^2),  \\
    \Delta\Gamma_2^{(12)}+\Delta\Gamma_2^{(21)} =&~  
    \frac{-m_{N_1}}{128\pi^2 m_{N_2}^3} \bigg[
    \big[ \Im(y^\nu_{2j} y^{\nu *}_{1j}\xi_{22}\xi_{12}^*)m_{N_1}
    +\Im(y^\nu_{2j} y^{\nu *}_{1j}\xi_{22}^*\xi_{12}) m_{N_2} \big]
    (m_{N_1}^2+m_{N_2}^2)   \nn \\
    & \quad
    +2\big[\Im(y^\nu_{2j} y^{\nu *}_{1j}\xi_{22}\xi_{12})
    m_{N_1} +\Im(y^\nu_{2j} y^{\nu *}_{1j}\xi_{22}^*\xi_{12}^*)
    m_{N_2}\big]m_{N_1}m_{N_2}  \,\bigg],  \\
    \Delta\Gamma_2^{(22)} =&~  
    \frac{\gamma_{N_2}}{128\pi^3} y^\nu_{2j} y^{\nu *}_{2j}\Im (\xi^2_{22})
    \ln \Bigl( \frac{m_{N_2}}{m_\chi} \Bigr) .
\end{align}

\bigskip

\section{Boltzmann equations and asymmetry formulae}
\label{sec:boltzmann-DeltaL}

\subsection{Boltzmann equations}
\label{sec:app-boltzmann}

\begin{figure}[t]
	\centering
	\includegraphics[height=0.2\textwidth]{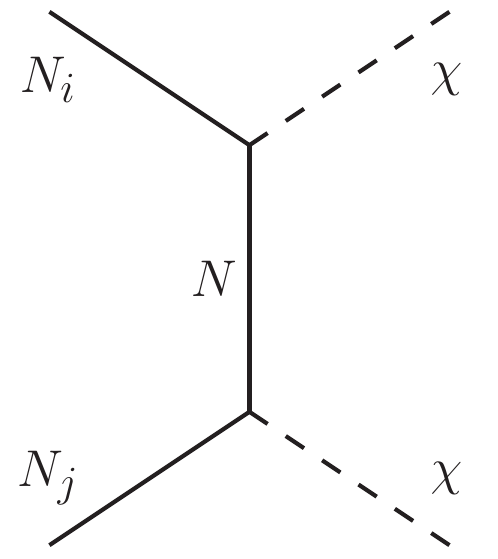}
	\qquad
	\includegraphics[height=0.2\textwidth]{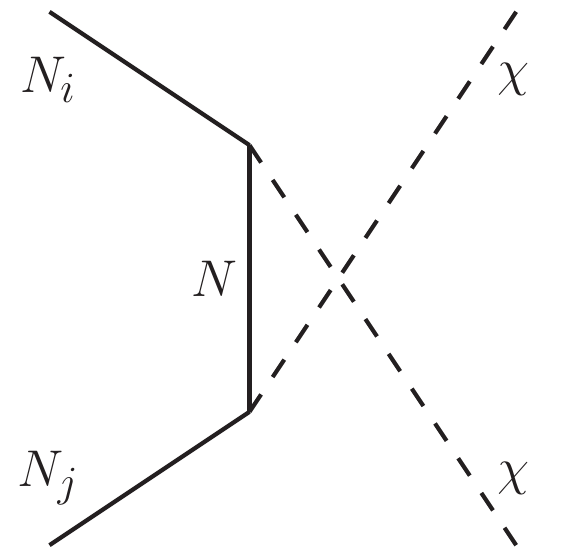}
	\qquad
	\includegraphics[height=0.2\textwidth]{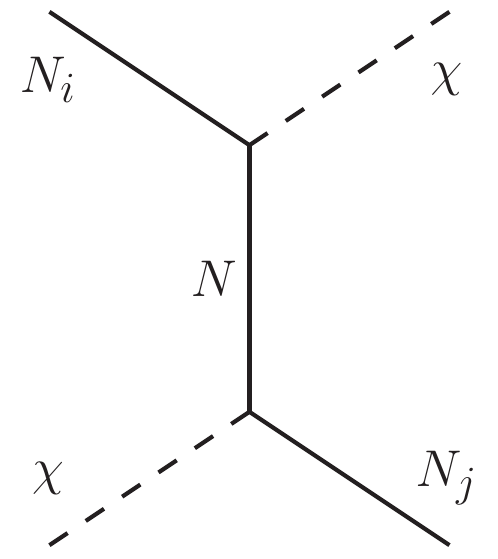}
	\caption{
		    The scattering processes with the new scalar coupling $\xi$.
	}
	\bigskip
	\label{fig:ScatteringFigures}
\end{figure}

The Boltzmann equations for the system of Eq.~\eqref{eq:Lagrangian} are given by

\begin{align}
 H x \frac{d Y_{N_1}}{dx} =&\;
 \frac{K_1 (m_{N_1}/T)}{K_2 (m_{N_1} /T)} Y_{N_1}^\eq \biggl[\sum_j \tilde\Gamma_{1j} \biggl( 1 - \frac{Y_{N_1}}{Y_{N_1}^\eq} \biggr)
 - \frac{1}{2} \sum_j \tilde\Gamma_{1j} \tilde\epsilon_{1j} \frac{Y_{\Delta L_j}}{Y^\eq_{\LL}}
 \nn \\
 &\; + \sum_j \Gamma_{1j} \biggl( \frac{Y_\chi}{Y^\eq_\chi} - \frac{Y_{N_1}}{Y^\eq_{N_1}} \biggr)
 - \frac{1}{2} \sum_j \Gamma_{1j} \epsilon_{1j} \frac{Y_\chi}{Y^\eq_\chi} \frac{Y_{\Delta L_j}}{Y^\eq_{\LL}}
 \biggr]
 \nn \\
 &\; + \frac{K_1(m_{N_2} /T)}{K_2 (m_{N_2} /T)} Y^\eq_{N_2} \Gamma_{N_2 \to N_1 \chi} \biggl( \frac{Y_{N_2}}{Y^\eq_{N_2}} - \frac{Y_{N_1}}{Y^\eq_{N_1}} \frac{Y_\chi}{Y^\eq_\chi} \biggr)
 + C_\scat,
 \label{eq:BE-N_1}
 \\
 H x \frac{d Y_{N_2}}{dx} =&\;
 \frac{K_1 (m_{N_2} /T)}{K_2(m_{N_2}/T)} Y^\eq_{N_2} \biggl[ \sum_j \tilde\Gamma_{2j} \biggl( 1 - \frac{Y_{N_2}}{Y^\eq_{N_2}} \biggr)
 - \frac{1}{2} \sum_j \tilde\Gamma_{2j} \tilde\epsilon_{2j} \frac{Y_{\Delta L_j}}{Y^\eq_{\LL}}
 \nn \\
 &\; + \sum_j \Gamma_{2j} \biggl( \frac{Y_\chi}{Y^\eq_\chi} - \frac{Y_{N_2}}{Y^\eq_{N_2}} \biggr)
 - \frac{1}{2} \sum_j \Gamma_{2j} \epsilon_{2j} \frac{Y_\chi}{Y^\eq_\chi} \frac{Y_{\Delta L_j}}{Y^\eq_{\LL}}
 \nn \\
 &\; - \Gamma_{N_2 \to N_1 \chi} \biggl( \frac{Y_{N_2}}{Y^\eq_{N_2}} - \frac{Y_{N_1}}{Y^\eq_{N_1}} \frac{Y_\chi}{Y^\eq_\chi} \biggr) \biggr]
 + C_\scat,
 \label{eq:BE-N_2}
 \\
 H x \frac{d Y_{\Delta L_i}}{dx} = &\;
 \sum_j \frac{K_1 (m_{N_j}/T)}{K_2(m_{N_j}/T)} Y^\eq_{N_j} \Biggl[\tilde\Gamma_{ji} \tilde\epsilon_{ji} \biggl( \frac{Y_{N_j}}{Y^\eq_{N_j}} - 1 \biggr)
 - \frac{1}{2} \tilde\Gamma_{ji} \frac{Y_{\Delta L_i}}{Y^\eq_{\LL}}
 \nn \\
 &\; + \Gamma_{ji} \epsilon_{ji} \biggl( \frac{Y_{N_j}}{Y^\eq_{N_j}} - \frac{Y_\chi}{Y^\eq_\chi} \biggr)
 - \frac{1}{2} \Gamma_{ji} \frac{Y_\chi}{Y^\eq_\chi} \frac{Y_{\Delta L_j}}{Y^\eq_{\LL}} \Biggr]
 \nn \\
 &\; + C_\scat,
 \label{eq:BE-DeltaL}
 \\
 H x \frac{dY_\chi}{dx} =&\;
 \sum_{i,j}  \frac{K_1 (m_{N_i}/T)}{K_2(m_{N_i}/T)} Y^\eq_{N_i}\biggl[
 \Gamma_{ij} \biggl( \frac{Y_{N_i}}{Y^\eq_{N_i}} - \frac{Y_\chi}{Y^\eq_{\chi}} \biggr)
 + \frac{1}{2} \Gamma_{ij} \epsilon_{ij} \frac{Y_\chi}{Y^\eq_{\chi}} \frac{Y_{\Delta L}}{Y^\eq_{\LL}} \biggr]
 \nn \\
 &\; + \frac{K_1 (m_{N_2}/T)}{K_2 (m_{N_2}/T)} Y^\eq_{N_2} \Gamma_{N_2 \to N_1 \chi} \biggl( \frac{Y_{N_2}}{Y^\eq_{N_2}} - \frac{Y_{N_1}}{Y^\eq_{N_1}} \frac{Y_\chi}{Y^\eq_{\chi}} \biggr)
 + C_\scat.
 \label{eq:BE-chi}
\end{align}
$C_\scat$ denotes the collision terms of the scattering divided by the entropy density with the on-shell contribution removed.
The NGB property of $\chi$ would generally suppresses the $s$-channel contribution of heavy modes, and the dominant one with the scalar coupling $\xi$ comes from the scattering processes 
$N N \leftrightarrow \chi \chi$ and 
$N \chi \leftrightarrow N \chi$ shown in Fig.~\ref{fig:ScatteringFigures}.
In deriving these equations,
the SM particles, in particular the leptons and the Higgs bosons, are assumed to be in thermal equilibrium:
\begin{align}
 Y_{L_j} = Y^\eq_\LL + \frac{1}{2} Y_{\Delta L_j},
 \quad
 Y_{\bar{L}_j} = Y^\eq_\LL - \frac{1}{2} Y_{\Delta L_j},
 \quad
 Y_{H} = Y_{\bar{H}} = Y^\eq_{H},
\end{align}
with the lepton asymmetry of $j$-th generation $Y_{\Delta L_j}$.
In this paper,
we analyze the Boltzmann equations with the single-flavored approximation.
The total lepton asymmetry is defined by
\begin{align}
 Y_{\Delta L} \coloneqq \sum_i Y_{\Delta L_i},
\end{align}
and the collision terms are approximated as in the following forms:
\begin{align}
 &\sum_i \Gamma_{ji} \frac{Y_{\Delta L_i}}{Y^\eq_{\LL_i}}
 \sim \Gamma_j \frac{Y_{\Delta L}}{Y^\eq_\LL},
 \quad
 \sum_{i} \Gamma_{ji} \epsilon_{ji} \frac{Y_{\Delta L_i}}{Y^\eq_{\LL_i}}
 \sim \Gamma_j \epsilon_j \frac{Y_{\Delta L}}{Y^\eq_\LL},
 \\
 &\sum_i \tilde{\Gamma}_{ji} \frac{Y_{\Delta L_i}}{Y^\eq_{\LL_i}}
 \sim \tilde{\Gamma}_j \frac{Y_{\Delta L}}{Y^\eq_\LL},
 \quad
 \sum_{i} \tilde{\Gamma}_{ji} \tilde{\epsilon}_{ji} \frac{Y_{\Delta L_i}}{Y^\eq_{\LL_i}}
 \sim \tilde{\Gamma}_j \tilde{\epsilon}_j \frac{Y_{\Delta L}}{Y^\eq_\LL}.
\end{align}
Using these equations,
the Boltzmann equations \eqref{eq:BE-N_1}--\eqref{eq:BE-chi} are rewritten as
\begin{align}
 H x \frac{dY_{N_1}}{dx} \approx &\;
 \frac{K_1 (m_{N_1}/T)}{K_2( m_{N_1}/T)} Y^\eq_{N_1}  \Biggl[ \tilde\Gamma_1 \biggl( 1 - \frac{Y_{N_1}}{Y^\eq_{N_1}} \biggr)
 - \frac{1}{2} \tilde\Gamma_1 \tilde\epsilon_1 \frac{Y_{\Delta L}}{Y^\eq_\LL}
 + \Gamma_1 \biggl( - \frac{Y_{N_1}}{Y^\eq_{N_1}} + \frac{Y_\chi}{Y^\eq_\chi} \biggr)
 \nn \\
 &\; - \frac{1}{2} \Gamma_1 \epsilon_1 \frac{Y_\chi}{Y^\eq_\chi} \frac{Y_{\Delta L}}{Y^\eq_\LL} \Biggr]
 + \frac{K_1 (m_{N_2} /T)}{K_2(m_{N_2}/ T)} Y^\eq_{N_2} \Gamma_{N_2 \to N_1 \chi} \biggl( \frac{Y_{N_2}}{Y_{N_2}^\eq} - \frac{Y_{N_1}}{Y^\eq_{N_1}} \frac{Y_\chi}{Y^\eq_{\chi}} \biggr)
 + C_\scat,
 \tag{\ref{eq:BE-N_1-sfa}}
 \\
 H x \frac{ d Y_{N_2}}{dx} \approx &\;
 \frac{K_1(m_{N_2}/T)}{K_2(m_{N_2}/T)} Y^\eq_{N_2} \Biggl[ \tilde\Gamma_2 \biggl( 1 - \frac{Y_{N_2}}{Y^\eq_{N_2}} \biggr)
 - \frac{1}{2} \tilde\Gamma_2 \tilde\epsilon_2 \frac{Y_{\Delta L}}{Y^\eq_\LL}
 + \Gamma_2 \biggl( - \frac{Y_{N_2}}{Y^\eq_{N_2}} + \frac{Y_\chi}{Y^\eq_\chi} \biggr)
 \nn \\
 &\; - \frac{1}{2} \Gamma_2 \epsilon_2 \frac{Y_\chi}{Y^\eq_\chi} \frac{Y_{\Delta L}}{Y^\eq_\LL}
 - \Gamma_{N_2 \to N_1 \chi} \biggl( \frac{Y_{N_2}}{Y^\eq_{N_2}} - \frac{Y_{N_1}}{Y^\eq_{N_1}} \frac{Y_\chi}{Y_\chi^\eq} \biggr) \Biggr]
 + C_\scat,
 \tag{\ref{eq:BE-N_2-sfa}}
 \\
 H x \frac{d Y_{\Delta L}}{dx} \approx &\;
 \sum_i
 \frac{K_1 (m_{N_i}/T)}{K_2 (m_{N_i}/T)} Y^\eq_{N_i}
 \Biggl[ \tilde\Gamma_i \tilde\epsilon_i \biggl( \frac{Y_{N_i}}{Y^\eq_{N_i}} - 1 \biggr)
 - \frac{1}{2} \tilde\Gamma_i \frac{Y_{\Delta L}}{Y^\eq_\LL}
 \nn \\
 &\; + \Gamma_i \epsilon_i \biggl( \frac{Y_{N_i}}{Y^\eq_{N_i}} - \frac{Y_\chi}{Y^\eq_\chi} \biggr)
 - \frac{1}{2} \Gamma_i \frac{Y_\chi}{Y^\eq_\chi} \frac{Y_{\Delta L}}{Y^\eq_\LL}
 \Biggr]
 + C_\scat,
 \tag{\ref{eq:BE-DeltaL-sfa}}
 \\
 H x \frac{d Y_\chi}{dx} \approx &\;
 \sum_i \frac{K_1 (m_{N_i}/T)}{K_2(m_{N_i}/T)} Y^\eq_{N_i} \Biggl[
 \Gamma_i \biggl( \frac{Y_{N_i}}{Y^\eq_{N_i}} - \frac{Y_\chi}{Y_\chi^\eq} \biggr)
 + \frac{1}{2} \Gamma_i \epsilon_i \frac{Y_\chi}{Y_\chi^\eq} \frac{Y_{\Delta L}}{Y^\eq_\LL} \Biggr]
 \nn \\
 &\; + Y^\eq_{N_2} \frac{K_1 (m_{N_2}/T)}{K_2 (m_{N_2}/T)} \Gamma_{N_2 \to N_1 \chi} \biggl( \frac{Y_{N_2}}{Y^\eq_{N_2}} - \frac{Y_{N_1}}{Y^\eq_{N_1}} \frac{Y_\chi}{Y^\eq_\chi} \biggr)
 + C_\scat.
 \tag{\ref{eq:BE-chi-sfa}}
\end{align}

\subsection{Phenomenological formulae for lepton asymmetry}
\label{sec:formula}

The Boltzmann equation of the lepton asymmetry (e.g. Eq.~\eqref{eq:BE-DeltaL-sfa}) has following form:
\begin{align}
 \frac{d Y_{\Delta L}(x)}{dx} = \mathcal{F}(x) - W(x) Y_{\Delta L}(x),
\end{align}
where $\mathcal{F}(x)$ is the $Y_{\Delta L}$ independent function describing the asymmetry production from decays and scatterings of $N_i$
and $W(x)$ is the washout function.
As discussed in Refs.~\cite{Barbieri:1999ma,Buchmuller:2004tu,Davidson:2008bu},
the solution of this equation can be expressed by
\begin{align}
 Y_{\Delta L}(x) = \int_{x_i}^{x} dx'\, \mathcal{F}(x')
 \exp \biggl[ - \int_{x'}^{x} dx''\, W(x'') \biggr]
 + Y_{\Delta L}(x_i) \exp \biggl[ - \int_{x_i}^{x} dx''\, W(x'') \biggr].
 \label{eq:washout-formal-sol}
\end{align}
Intuitively,
the integral of $\mathcal{F}$ means the freeze-in like production \cite{Hall:2009bx}
and $e^{-\int dx' W(x')}$ denotes the washout suppression.
Using this equation,
we will show the approximated values of the lepton asymmetry,
$\mathcal{Y}^{\mathrm{FI}}_{\Delta L}(z_1)$ and $\mathcal{Y}^{\mathrm{WO}}_{\Delta L}(\infty)$ in the following part of this section.

\subsubsection{Case for freeze-in from $N_2$}

$N_2$ is mainly produced by the first term of the RHS in Eq.~\eqref{eq:BE-N_2-sfa}.
Using the approximation for the modified Bessel function
\begin{align}
 K_n(x) \underset{x \sim 0}{\sim} \frac{(n-1)!}{2} \biggl( \frac{x}{2} \biggr)^{-n},
\end{align}
the Boltzmann equation is approximately written as
\begin{align}
 \frac{d Y_{N_2}}{dx} \approx \frac{1}{H x} \frac{K_1 \left( \frac{m_{N_2}}{m_{N_1}} x \right)}{K_2 \left( \frac{m_{N_2}}{m_{N_1}} x \right)}
 \tilde{\Gamma}_2 Y^\eq_{N_2}
 \sim \frac{135 \sqrt{10}}{2 \pi^5 g_*^{1/2} g_*^S}
 \frac{M_P m_{N_2} \tilde{\Gamma}_2}{m_{N_1}^3} x^2,
\end{align}
and the yield of $N_2$ is scaled by
\begin{align}
 \bm{Y}_{N_2} = \frac{45 \sqrt{10}}{2 \pi^5 g^{1/2}_* g^S_*}
 \frac{M_P m_{N_2} \tilde{\Gamma}_2}{m_{N_1}^3} x^3.
\end{align}
This $N_2$ produces $\chi$ and $N_1$ via $N_2 \to N_1 \chi$ process,
and their yields are also written as
\begin{align}
 \bm{Y}_\chi &= \frac{135}{2 \pi^6 g_* g_*^S}
 \frac{M_P^2 m_{N_2}^2 \tilde{\Gamma}_2 ( \Gamma_{N_2 \to N_1 \chi} + \Gamma_2)}{m_{N_1}^6} x^6,
 \\
 \bm{Y}_{N_1} &= \frac{45 \sqrt{10}}{2 \pi^5 g^{1/2}_* g^S_*}
 \frac{M_P \tilde{\Gamma}_1}{m_{N_1}^2} x^3
 + \frac{135}{2 \pi^6 g_* g_*^S}
 \frac{M_P^2 m_{N_2}^2 \tilde{\Gamma}_2 ( \Gamma_{N_2 \to N_1 \chi} + \Gamma_2)}{m_{N_1}^6} x^6.
\end{align}
where the contribution of the first term of the RHS in Eq.~\eqref{eq:BE-N_1-sfa} is included, which is scaled by $x^3$ as in the same way to $N_2$.
Using these functions,
we will evaluate the lepton asymmetry.
From Eq.~\eqref{eq:BE-DeltaL-sfa},
The washout function
\begin{align}
 W_2 \coloneqq \frac{1}{2 H x} \frac{K_1 ( m_{N_2} x / m_{N_1})}{K_2 ( m_{N_2} x / m_{N_1})}
 \biggl( \tilde{\Gamma}_2 + \Gamma_2 \frac{\bm{Y}_\chi}{Y^\eq_\chi} \biggr)
 \frac{Y^\eq_{N_2}}{Y^\eq_\LL}
 + \frac{1}{2 H x} \frac{K_1 (x)}{K_2(x)}
 \biggl( \tilde{\Gamma}_1 + \Gamma_1 \frac{\bm{Y}_\chi}{Y^\eq_\chi} \biggr)
 \frac{Y^\eq_{N_2}}{Y^\eq_\LL},
 \label{eq:washout-function-W2}
\end{align}
is read,
and the yield is evaluated by using \eqref{eq:washout-formal-sol} as
\begin{align}
 \mathcal{Y}_{\Delta L}^{\mathrm{FI}} (x) = \int_{0}^{x} dx' \, \mathcal{F}_2 (x')
 \exp \biggl[ - \int_{x'}^{x} dx''\, W_2(x'') \biggr],
\end{align}
with
\begin{align}
 \mathcal{F}_2 =&\;
 \frac{1}{H x}  \frac{K_1 (m_{N_2}x /m_{N_1})}{K_2(m_{N_2}x/ m_{N_1})}
 \Gamma_2 \epsilon_2 \biggl( \bm{Y}_{N_2} - \frac{Y^\eq_{N_2}}{Y^\eq_\chi} \bm{Y}_\chi \biggr)
 + \frac{1}{H x} \frac{K_1(x)}{K_2(x)} \Gamma_1 \epsilon_1
 \biggl( \bm{Y}_{N_1} - \frac{Y^\eq_{N_1}}{Y^\eq_\chi} \bm{Y}_\chi \biggr).
\end{align}
If the washout suppression is weak enough,
which is the case of the neutrino Yukawa couplings being small,
the lepton asymmetry is mainly produced by the RH neutrinos before getting into the thermal bath
and the total amount is evaluated as
\begin{align}
 Y_{\Delta L} \approx \mathcal{Y}^{\mathrm{FI}}_{\Delta L} (z_1),
\end{align}
where
\begin{align}
 z_1 = \mathop{\mathrm{min}} ( z_{N_2}, z_\chi),
 \quad
 \bm{Y}_{N_2} (z_{N_2}) = \frac{45}{\pi^4 g^S_*},
 \quad
 \bm{Y}_\chi(z_\chi) = \frac{45}{2\pi^4 g^S_*}.
\end{align}
This is regarded as the asymmetry by the freeze-in production including the weak washout effects.

\begin{figure}[t]
	\centering
	\includegraphics[width=0.45\textwidth]{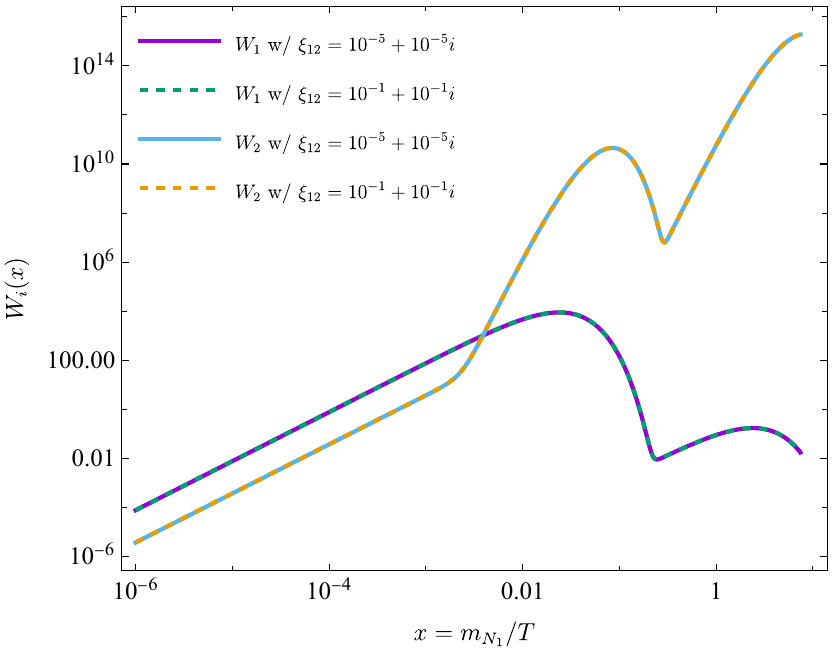}
	\qquad
	\includegraphics[width=0.45\textwidth]{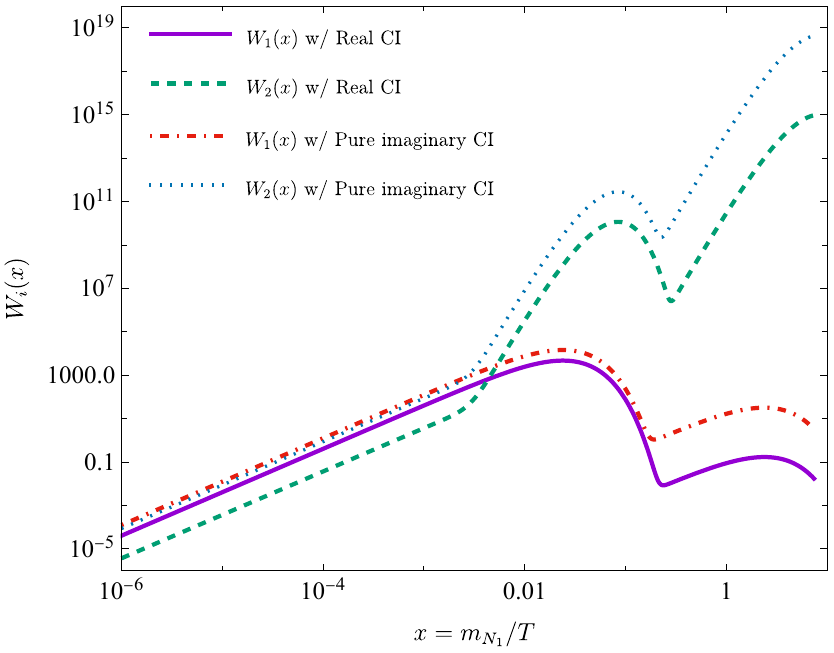}
	\caption{
	The time evolution of the washout functions $W_{1,2}$ [\eqref{eq:washout-function-W2} and \eqref{eq:washout-function-W1}].
	(Left) The $\xi$ dependence of $W_{1,2}$ with the real CI parameters \eqref{eq:Real-CI}.
	(Right) The Yukawa coupling dependence of $W_{1,2}$.
	}
	\label{fig:washout-function-compare}
\end{figure}

\subsubsection{Case for strong $N_1$ washout}

If the washout of $N_1$ is strong,
the lepton asymmetry is mainly determined by this,
and the contribution from $\mathcal{Y}^{\mathrm{FI}}_{\Delta L}(z_1)$ is exponentially suppressed.
The washout function of this case is given by
\begin{align}
 W_1 \coloneqq \frac{1}{2 H x} \frac{K_1 (m_{N_2} x /m_{N_1})}{K_2 (m_{N_2} x / m_{N_1})} ( \tilde{\Gamma}_2 + \Gamma_2 ) \frac{Y^\eq_{N_2}}{Y^\eq_\LL}
 + \frac{1}{2 H x} \frac{K_1 (x)}{K_2(x)} ( \tilde{\Gamma}_1 + \Gamma_1) \frac{Y^\eq_{N_1}}{Y^\eq_\LL},
 \label{eq:washout-function-W1}
\end{align}
where $Y_\chi \approx Y^\eq_\chi$ is assumed.
Then, the lepton asymmetry is written as
\begin{align}
 \mathcal{Y}^{\mathrm{WO}}_{\Delta L } = \int_0^x dx'\, \mathcal{F}_1(x')
 \exp \biggl[ - \int_{x'}^{x} dx'' \, W_1(x'') \biggr],
\end{align}
where we introduce the following function
\begin{align}
 \mathcal{F}_1\coloneqq&
 \frac{1}{H x} \frac{K_1(m_{N_2} x/ m_{N_1})}{K_2( m_{N_2} x /m_{N_1})}
 \Gamma_2 \epsilon_2 \biggl( \frac{ \tilde{\Gamma}_2 + \Gamma_2}{m_{N_2}} \biggr) Y^\eq_{N_2}
 + \frac{1}{H x} \frac{K_1(x)}{K_2(x)}
 \Gamma_1 \epsilon_1 \biggl( \frac{\tilde{\Gamma}_1 + \Gamma_1}{m_{N_1}} \biggr) Y^\eq_{N_1}.
\end{align}
Using these functions,
the lepton asymmetry is given by
\begin{align}
 Y_{\Delta L} \approx \mathcal{Y}^{\mathrm{WO}}_{\Delta L } (\infty).
\end{align}

\subsubsection{Washout functions}

The time dependence of the washout functions \eqref{eq:washout-function-W2} and \eqref{eq:washout-function-W1} are shown in Fig.~\ref{fig:washout-function-compare}.
For the neutrino Yukawa couplings, the real CI parameters \eqref{eq:Real-CI} and pure imaginary CI parameters \eqref{eq:Pure-imaginary-CI} are used.
We find in the left panel these washout functions little depend on the complex coupling $\xi$ as the neutrino Yukawa couplings give the dominant effects to the washout. 
The right panel shows the Yukawa coupling dependence of the washout functions.
Larger Yukawa couplings give larger washout functions, which means the lepton asymmetry is strongly washed out due to the factor $\exp(-W_i)$.

\bigskip

\newcommand{\arxivfont}{\rmfamily}
\bibliographystyle{yautphys}
\bibliography{ref}

\end{document}